# Exo–Zodiacal Dust Levels for Nearby Main–Sequence Stars: A Survey with the Keck Interferometer Nuller

R. Millan-Gabet

California Institute of Technology, NASA Exoplanet Science Institute, Pasadena, CA 91125, USA

R.Millan-Gabet@caltech.edu

E. Serabyn, B. Mennesson, W. A. Traub

Jet Propulsion Laboratory, California Institute of Technology, 4800 Oak Grove Drive, Pasadena, CA 91109, USA

R. K. Barry, W. C. Danchi, M. Kuchner

NASA Goddard Space Flight Center, Exoplanets and Stellar Astrophysics Laboratory, Code 667, Greenbelt, MD 20771, USA

S. Ragland, M. Hrynevych, J. Woillez

Keck Observatory, 65-1120 Mamalahoa Hwy, Kamuela, HI 96743, USA

and

K. Stapelfeldt, G. Bryden, M. M. Colavita, A. J. Booth

Jet Propulsion Laboratory, California Institute of Technology, 4800 Oak Grove Drive, Pasadena, CA 91109, USA

## ABSTRACT

The Keck Interferometer Nuller (KIN) was used to survey 25 nearby main sequence stars in the mid–infrared, in order to assess the prevalence of warm circumstellar (exozodiacal) dust around nearby solar–type stars. The KIN measures circumstellar emission by spatially blocking the star but transmitting the circumstellar flux in a region typically $0.1 - 4$ AU from the star. We find one significant detection ($\eta$ *Crv*), two marginal detections ($\gamma$ Oph and $\alpha$ *Aql*), and 22 clear non–detections. Using a model of our own Solar System's zodiacal cloud, scaled to the luminosity of each target star, we estimate the equivalent number of target zodis needed to match our observations. Our three zodi detections are $\eta$ Crv ($1250 \pm 260$), $\gamma$ Oph ($200 \pm 80$) and $\alpha$ Aql ($600 \pm 200$), where the uncertainties are $1\sigma$. The 22 non–detected targets have an ensemble weighted average consistent with zero, with an average individual uncertainty of 160 zodis ($1\sigma$). These measurements represent the best limits to date on exozodi levels for a sample of nearby main sequence stars. A statistical analysis of the population of 23 stars not previously known to contain circumstellar dust (excluding $\eta$ Crv and $\gamma$ Oph) suggests that, if the measurement errors are uncorrelated (for which we provide evidence) and if these 23 stars are representative of a single class with respect to the level of exozodi brightness, the mean exozodi level for the class is $< 150$ zodis ($3\sigma$ upper–limit, corresponding to 99% confidence under the additional assumption that the measurement errors are Gaussian). We also demonstrate that this conclusion is largely independent of the shape and mean level of the (unknown) true underlying exozodi distribution.

## 1. Introduction

The present day Solar System contains interplanetary dust. The term *zodiacal* usually refers to the dust present in the inner Solar System, out to $\sim 5$ AU. It currently has a fractional luminosity of $L_{dust}/L_{\odot} \sim 10^{-7}$ (Backman & Paresce 1993) and $10^{-10}$ the mass of the planets, but an infrared luminosity 100× larger. Because the survival times for small particles in the radiation and wind environment of the star is a few 100 to a few 1000 yrs, any such circumstellar dust observed in stars older than a few tens of millions of years must be recently formed and continuously generated. The presence of prominent dust bands in the solar zodiacal cloud associated with asteroid families suggest that the zodiacal cloud arises from the breakup of main belt asteroids (see e.g. Dermott et al. 1984). However, recent models of zodiacal dust production imply that the splitting of short period comets could be the dominant source (Nesvorny et al. 2010).

Circumstellar dust around other mature stars was originally discovered, unexpectedly, by IRAS (Aumann et al. (1984), see e.g. the review by Backman & Paresce (1993)), and has since been observationally studied by a variety of ground and space observatories, via both their thermal and scattered emission. Most commonly, the phenomenon reveals itself as long–wavelength fluxes in excess of what is expected from the stellar photosphere alone, but spatially resolved images of the outer disk regions have also been obtained for a few of the most extreme systems (see e.g. the review by Zuckerman 2001). The term *debris* disk usually refers to the entire dust distribution, extending to 100s of astronomical units (AU) from the Sun.

The presence of high levels of cold outer dust around main sequence stars is now known to be a ubiquitous phenomenon (e.g. 30% of all A–stars, and 13% of solar–type stars Rieke et al. 2005; Su et al. 2006; Bryden et al. 2006; Beichman et al. 2006b). However, very few stars have had positive detections of excess flux at wavelengths $< 30\mu$m (Beichman et al. 2006a; Lawler et al. 2009). Although this rarity appears to be consistent with evolution models and detection thresholds, the fact remains that much less is known about levels of inner warm dust, of most interest to extra–solar terrestrial planets searches. The measurement is difficult, because the dust emission is close to and faint compared to the parent star. Thus, we currently have no firm estimates of the warm zodi brightness around nearby stars, or around any stars. Interestingly, the few systems that have been imaged in some detail at sub–mm wavelengths display a striking variety of complex morphological features. This forces caution when interpreting spatially unresolved observations, and highlights the difficulties in attempting to infer levels of inner warm dust from measurements made at wavelengths which probe very different spatial scales.

Searches for exozodiacal emission from warm inner dust have been attempted from the ground (e.g. Kuchner et al. 1998; Liu et al. 2004) and space (e.g. Beichman et al. 2006a; Lawler et al. 2009). Due to current limitations of the observational techniques, known exozodi disks have much higher dust densities than the Solar System (e.g. $L_{dust}/L_\star > 10^{-4}$ from the Lawler et al. (2009) survey); the present–day Solar System zodiacal levels would be currently undetectable around other stars.

Studying exozodi clouds is of interest for a variety of reasons. Among them, the time–scales for debris disk evolution may help understand terrestrial planet formation (see e.g. Wyatt 2008), and disk structure may be used to infer the presence of perturbing unseen exo–planets (Wolf et al. 2007; Stark & Kuchner 2008), examples of which have now been directly imaged (Marois et al. 2008; Kalas et al. 2008). Moreover, both the levels of exozodi emission and their spatial structure act as sources of noise that may hinder the direct detection and characterization of terrestrial exoplanets by spaced–based coronographic or interferometric techniques (e.g. Beckwith 2008). Indeed the largest uncertainty in estimating planet detection efficiencies is due to exozodi dust. Since exozodi photons impact the required integration times and sample sizes for a given mission lifetime, knowledge of exozodi levels and structure for all candidate stars would allow a greatly optimized instrument and observing strategy design.

Mid–infrared interferometry has provided exozodi measurements at high spatial resolution for specific known high–dust stars (Stark et al. 2009; Smith et al. 2009). Interferometric techniques have also enabled the identification of an intriguing source of near–infrared excess around some main sequence stars (Absil et al. 2008, 2006; Akeson et al. 2009; Absil et al. 2009). Obtaining exozodi measurements for relatively large samples of representative nearby main sequence stars was the primary scientific driver for the development of the Keck Interferometer Nuller (KIN), and for the execution of a 1–year long intensive Key Science observing campaign shared among three selected proposals. This paper summarizes the results from one of three Key Science programs selected ("The Keck Interferometer Nuller Survey of Exozodiacal Dust around Nearby Stars ", Principal Investigator: G. Serabyn).

## 2. The Sample

Of all nearby stars, $\sim$ 85 dwarfs and sub–dwarfs fall within the sensitivity and observability limits of the KIN. Given the expected time allocation for the Key Science programs, we down–selected this list to 40 high priority objects, containing systems both with and without known debris disk emission (a.k.a "high dust" and "low dust", respectively) as inferred from mid–infrared spectrophotometric measurements. The selecting committee ultimately assigned 5 high dust and 24 low dust objects to the program described here. Of those, 25 systems were actually observed, 2 high dust ($\eta$ Crv and $\gamma$ Oph) and 23 low dust. Observations took place in service mode during the period February 2008 – January 2009, over 32 nights shared with the other two Key Science programs. Table 1 describes our list of observed targets, including all the brightnesses relevant to the various KIN subsystems, and the stellar parameters relevant to our modeling approach.

## 3. The Keck Interferometer Nuller

### 3.1. Instrument Overview

The Keck Interferometer Nuller (Colavita et al. 2009) operates in N–band ($8.0-13.0\mu$m) and combines the light from the two Keck telescopes as an interferometer with a physical baseline length $B \sim 85$ m. The KIN produces a dark fringe through the phase center ("Nulling"). The adjacent bright fringe (through which flux is transmitted), projects onto the sky at an angular separation $\lambda/2B = 10$ mas, or 0.1 AU at the median distance to the stars in our sample (10.5 pc), and for $\lambda = 8.5\,\mu$m (the effective wavelength of the KIN bandpass). Thus, the instrument is sensitive to circumstellar dust located as close to the central star as these spatial scales (i.e. "inner working angle"). Blackbody emission peaks at $8.5\,\mu$m for $T \sim 432$ K. For the median luminosity of the stars in our sample ($2.2L_{\odot}$), dust particles in thermal equilibrium at this temperature are located at a stellocentric radius of 0.3 AU. Therefore, the KIN fringe spacing matches well the expected location of relatively warm dust, located in the inner disk regions. The KIN can observe objects as faint as N(flux) = 1.7 Jy, as long as they also have Kmag < 6 (K–band co–phasing limit) and Jmag < 8.5 (angle tracking limit).[1]

The response of any interferometer may be understood by projecting the fringe pattern on the sky: what is measured is the astrophysical flux from the surface brightness transmitted through the fringe pattern. For the work presented here, we measure and

[1]see *http://nexsci.caltech.edu/software/KISupport/nulling/* for a full description of the instrument parameters.

calibrate the transmitted flux (expected to be small) due to dust surrounding the central target stars. Thus, we refer to the basic observable as the "flux leakage" or simply the "leak" (the inverse quantity, the null depth, is an equivalent and also frequently used term, i.e. leak = 0.01 implies null depth 100:1). The amount of flux leakage not attributable to the finite size of the central stars is referred to as "excess leakage", and by measuring it we can learn about the amounts of circumstellar dust present.

In order to achieve good accuracy of the calibrated leak measurements, the KIN utilizes an architecture in which each Keck pupil is split into two halves, resulting in two Keck–Keck long baselines, and two short baselines (4 m) formed between the two halves of each Keck telescope. In order to accommodate the large dynamic range between a star and any surrounding dust, the star is nulled on the Keck–Keck baselines. In order to detect small leakage signals in the presence of the large mid–infrared background, the nulled outputs are combined on the short baselines in a standard Michelson combiner (a.k.a. the "cross–combiner", or XC) with fast optical path difference modulation (i.e. "interferometric chopping", see also Mennesson et al. 2005). The output of the short baseline combiners (for which any object appears essentially unresolved) when the long baselines are "at peak" also provide the necessary flux normalization of the leak measurement. In essence then, the KIN measurement is the ratio of the amplitudes of the short baseline combiner fringes when the long baseline combiner is set at null, divided by the same quantity when the long baseline combiner is set at peak:

$$L_{raw} = \frac{\text{XC fringe amplitude at null}}{\text{XC fringe amplitude at peak}} \tag{1}$$

The many details involved in making this measurement are described in Colavita et al. (2009) and Colavita et al. (2010). As emphasized in those references, in a ground–based background limited environment, achieving a high level of suppression of the central star (i.e. deep nulls) is not the most important consideration. Equally important is being able to calibrate the leaks well; and the KIN four–beam architecture results in a better calibration of the measured leakages (or equivalent visibilities) compared to standard Michelson interferometry at mid–infrared wavelengths.

Due to various instrumental factors (diffraction, material absorption, pinhole mode matching), the KIN spectral responsivity is strongly peaked toward the blue end of the bandpass; the $8 - 9\mu$m bin contains most of the signal–to–noise ratio (SNR). Also, the red end of the spectrum is affected by poorer calibration quality (believed to be caused by residual correlation in the short baseline combiners arising from telescope thermal emission). Thus, for the analysis presented here, we only use the $8 - 9\,\mu$m spectral bin, most sensitive to exozodi detection.

### 3.2. Sky Response

As noted above, the KIN response may be understood as the flux from the astrophysical source that is transmitted through its fringe pattern projected on the sky. Due to its four–beam architecture, the KIN beam pattern consists of three terms:

1. Point spread function (PSF) of each Keck half–aperture ($T_{PSF}$); this is well approximated at $8.5\,\mu$m by an elliptical Gaussian of FWHM = 490 × 440 mas. For the observations presented here, the Keck rotator angles are oriented such that this pattern has the major–axis along the East–West direction, and rotates as the target moves across the sky such that the minor–axis always points toward North.

2. Fringes of the short baseline combiner ($T_{XC}$):

$$T_{XC} = \cos(2\pi(x \cdot u_{xc} + y \cdot v_{xc})) \tag{2}$$

where $(x, y)$ are coordinate offsets in right ascension and declination, and $(u_{xc}, v_{xc})$ are the corresponding short baseline spatial frequencies. These fringes are perpendicular to the half–pupil PSF major–axis, and rotate in the same way. Projected on the sky, this fringe pattern is relatively "broad": at 8.5 $\mu$m the fringe spacing for the physical 4 m short baseline is 440 mas.

3. Fringes of the long baseline combiner ($T_{long}$):

$$T_{long} = \frac{1}{2} \cdot (1 \mp \cos(2\pi(x \cdot u + y \cdot v)) \quad (3)$$

where $(u, v)$ are the long baseline spatial frequencies, and the $\mp$ corresponds to the null/peak configuration, respectively. This fringe pattern can have any orientation with respect to $T_{PSF}$ and $T_{XC}$. As mentioned above, in this "fine" fringe pattern the dark and bright fringes are spaced by an angular separation projected on the sky of 10 mas, at 8.5 $\mu$m and for the physical 85 m KI baseline.

The total KIN transmission pattern is thus:

$$T(x, y, u, v, u_{xc}, v_{xc}, \lambda) = T_{PSF} \cdot T_{XC} \cdot T_{long} \quad (4)$$

Through the dependence on the spatial frequencies, the instantaneous KIN pattern depends on wavelength (8.5 $\mu$m) and on the hour angle and declination of the target being observed. Figure 1 shows an example of the KIN pattern terms. The KIN is sensitive to circumstellar dust by its ability to measure its flux transmitted through the total KIN fringe pattern (center panel of Figure 1) i.e. dust located from $\sim$ 10 mas (0.1 AU at 10 pc) out to the $\sim 490 \times 440$ mas field–of–view. However, the short baseline fringes ($T_{XC}$) also act to limit the KIN's ability to detect outer dust. Indeed, in the example shown in Figure 1 (top–right panel), $T_{XC}$ goes to zero at $\sim 110$ mas along the small baseline direction, so that the "effective FOV" is only $\sim 110 \times 440$ mas ($\sim 1.1 \times 4.4$ AU at 10 pc).

For an object with a brightness distribution $I(x, y, \lambda)$, the expected monochromatic leak is thus:

$$L_{calculated}(u, v, u_{xc}, v_{xc}, \lambda) = \frac{\int\int I(x, y, \lambda) \cdot T(x, y, u, v, u_{xc}, v_{xc}, \lambda)^{null} dxdy}{\int\int I(x, y, \lambda) \cdot T(x, y, u, v, u_{xc}, v_{xc}, \lambda)^{peak} dxdy} \tag{5}$$

where the double integral is performed over the KIN field–of–view, set by the the half–pupil PSF, as described above.

In order to provide some physical intuition, we note that for an object of angular size much smaller than the fringe spacing of the long baseline, the nulled signal would ideally be zero, resulting in $L = 0$. If, on the other hand, the object is large and its brightness distribution spans many long baseline fringes, the flux transmitted at null or at peak are similar and $L = 1.0$. Detailed descriptions of the theory behind the nuller measurement may also be found in Traub & Oppenheimer (2010) and Serabyn et al. (2011).

### 3.3. Data Reduction, Calibration and Errors

End–to–end data reduction and calibration was performed by the Keck Interferometer Project using their pipeline and external calibration package (*nullCalib*[2]). Here we summarize the steps involved and the various sources of measurement error.

During observing, a micro–sequence is executed during which the short baseline optical path modulation is always active (one fringe measurement every 25 msec), and the nullers alternate between the null and peak states (for 250 msec and 50 msec, respectively).

[2]http://nexsci.caltech.edu/software/V2calib/nullCalib/

Depending on the object brightness, this sequence is repeated for 10–15 min, thus many thousand independent estimates of the leak are formed for each observation. From the scatter of these measurements a “formal” error is estimated for the average leak measurement corresponding to each observation.

Following standard practice, in order to monitor and subtract the instrument’s transfer function (i.e. the non–zero leak that is measured when observing a point source located at the phase center), observations of targets of interest were interleaved with observations of calibrator stars of known N–band angular diameters. Thus, the calibrated leak is:

$$L_{calibrated} = (L_{raw}^{target} - L^{system}) \tag{6}$$

where $L^{system}$, at the time of the target observations, is obtained by interpolation from the net leak measurements of the bracketing calibrator stars, after subtracting the calibrator leak expected given its angular diameter. This calculation is described in more detail in Appendix C of Colavita et al. (2009), and all the steps are applied by the package *nullCalib*.

The required calibrator angular diameters were measured using the simultaneous K–band fringe tracker data, and converted to N–band diameters using standard limb–darkening relations. The procedure is described in detail in Appendix A of Colavita et al. (2009); for clarity we repeat here some of the most relevant aspects. The approach was to treat the nulling targets as calibrators, since at K–band they are expected to effectively be simple naked stars. Although some of the calibrators are small ($< 1$ mas) compared to the KI resolution, for such small calibrators the 20–30% precision obtained in those cases results in uncertainties on the calibrated leaks well below the leak measurement errors (best–case 0.2%, see below). At the other end of the size range, for a 3 mas calibrator, the largest in our sample, the diameter only needs to be known with 10% precision or better in order to not add significant error to the calibrated leaks. Thus, the calibration procedure is largely insensitive to resonable errors in the adopted calibrator angular diameters. The accuracy of our measured calibrator angular diameters was evaluated against estimates obtained using surface brightness relations; in all cases the discrepancies also result in a calibrated leakage differences smaller than the best–case measurement error. The uncertainties in the calibrator diameters are taken into account and propagated into the formal error of the calibrated leaks. The relevant parameters for the calibrators used for our sample are listed in Table 2.

We note that we make the assumption that there is no source of calibrator leak other than its angular extent, i.e. the calibrators have no excess N–band emission, which if unaccounted for would directly lead to underestimated exozodi emission levels. However, of 56 calibrators used in this study, one is a main–sequence star and all others are giants (i.e. none are super–giants), and all but three have spectral types K0–M0; both of which minimize the possibility of infrared emission above photospheric levels (e.g. Cohen et al. 1999, and references therein). Furthermore, all the calibrators used have been selected to have IRAS 12$\mu$m/25$\mu$m flux ratios that agree, within the photometric errors, with that of a blackbody of the calibrator effective temperature. These methods, however, do not insure that our calibrators are free from low level dust emission at N–band (few percent or less), undetected in the K–band diameter comparisons and IRAS flux ratios. But we can use our own data to place some limits on the possible level of exozodi emission around our calibrators. Indeed, if calibrator exozodi emission were high, and variable from star to star, this would be apparent in correspondingly large leak fluctuations among the different calibrators. Specifically, we have compared the variations in system leak estimates made when same calibrator is used on either side of the target observation (only instrument variations expected) with the same quantity computed when two different calibrators are used (variations due to instrument plus calibrator exozodi level). Within measurement errors, we see no difference between those two cases, implying that our calibrators do not contain large amounts of exozodi dust (in the language of Section 5, $\lesssim 100$ zodi impact on average on the exozodi levels derived for the target stars).

We also include "external" errors in the calibrated leaks, which have been estimated from the night–to–night repeatability of multiple sets of calibrated data taken on the same star (Colavita et al. 2009, 2010). In summary then, the errors on $L_{calibrated}$ contain two terms: (a) a formal error derived from the scatter of the leak measurements in each observation, and which also contains the uncertainties associated with the estimate of the system leak (mainly due to calibrator diameter uncertainties), and (b) the external error just described. Table 3 shows the measured calibrated leak and errors for each observation. As can be seen, for the stars in our sample the formal error is typically $\sigma^{formal}_{L_{calibrated}} = 0.001$ to 0.004. The external errors are wavelength and flux dependent; for the stars in our sample they are in the range: $\sigma^{ext.}_{L_{calibrated}} = 0.002$ to 0.0035.

Validation of the KIN response and calibration was evaluated by the project using a test system for which the expected leak could be calculated; in this case a binary system with a well known orbit (Appendix B of Colavita et al. 2009).

Table 3 also contains the calculated leak expected from the target star itself ($L_{\star}$), which is needed in order to derive the excess leak due to the exozodi cloud, as described below. The calibrated leak data are also shown in Figure 2. We emphasize that at this stage we do not average the multiple leak measurements that are available for a given target, because variations among leaks measured at various times may in principle contain a contribution from the changing projected baseline fringe spacing and orientation. Thus we first model our measurements for a specific exozodi model, and average the results after this conversion, as described in Section 4.4. Figure 3 also summarizes our measurements but only for the wideband ($8-9\,\mu$m) channel used in the analysis presented here.

## 4. Data Modeling

### 4.1. Modeling Exozodi Clouds

In order to interpret our measurements and compare them to the Solar System case, we use the *Zodipic* code[3] to create images of zodi clouds around each of our targets. *Zodipic* synthesizes brightness distributions of exozodiacal clouds based on the empirical fits to the observations of the solar zodiacal cloud made by COBE (Kelsall et al. 1998).

When *Zodipic* generates a model brightness distribution for a zodi disk analog around a star other than the Sun, the dust has the same optical depth at 1 AU and radial density profile as in the Solar System. As a convenient unit, we refer to this model as corresponding to "1–zodi", which we denote $z = 1$. *Zodipic* scales the radial temperature profile with stellar luminosity, and the inner dust radius is set by a dust sublimation temperature set at 1500 K (the dust inner radius is thus dependent on stellar spectral type, but $z = 1$ models around any star have a fractional dust luminosity $L_{dust}/L_\star \simeq 10^{-7}$, the Solar System value). In the *Zodipic* code the dust density can be treated as a free parameter, allowing to generate brightness distributions for scaled version of the Solar System (the total flux due to the circumstellar dust scales linearly with $z$). In the next section, we describe our procedure for converting the calibrated leaks to an exozodi dust density, parametrized in terms of a number ($z$) of zodis.

We note that the zodiacal models used here include only the smooth component of interest, i.e. the Earth trailing blob and asteroidal dust bands are not included. We also note that increasing the optical depth of the cloud increases the collision rate, which affects the cloud structure, a physical process which is not taken into account by *Zodipic*. In a zodiacal cloud, grain–grain collisions become important for grains above a critical size, $\sim 30 - 150\,\mu$m in the solar zodiacal cloud (Fixsen & Dwek 2002), and which scales inversely as the optical depth of the disk (Kuchner & Stark 2010). Since this critical grain size reaches $10\,\mu$m for a disk with about 3–15 zodis worth of dust, we expect that disks with more than roughly a few tens of zodis should begin to show morphological changes at KIN wavelengths because of the collision destruction of grains in the center of the disk. Our models, from *Zodipic*, are strictly linear in the dust density, and do not take this collisional depletion into account; this level of analysis is left to future studies.

---

[3]http://ssc.spitzer.caltech.edu/dataanalysistools/tools/contributed/general/zodipic/

We also note that our procedure assumes that the exo–zodi density in the inner–most regions of KIN sensitivity ($\sim$ 10 mas or 0.1 AU at 10 pc, as described above) follows the same radial density profile of the Kelsall et al. (1998) model, which was based on COBE/DIRBE measurements made at larger stellocentric radii, 0.9 AU or larger. However, measurements of the Solar zodiacal cloud made by the Helios probes as close as 0.3 AU (Leinert et al. 1981), and measurements in the solar F corona (MacQueen & Greeley 1995), both find radial density profiles with exponents of $\sim$ 1.3, in agreement with the Kelsall et al. (1998) model.

The *Zodipic* images generated are $512 \times 512$ pixels, with a scale of 2 mas/pixel, i.e. $10\times$ finer than the long baseline fringe spacing. The image size is thus $1024 \times 1024$ mas, a good match to the KIN field–of–view (given by $T_{PSF}$). Figure 1 includes an example *Zodipic* image for a Solar System analog at 10 pc. We note that this image size (in pixels) keeps the computation times short, but at this spatial resolution the stellar disk would not be well sampled. Therefore, the *Zodipic* images generated do not include the central stars, they represent only the exozodi brightness distribution ($I_{zodi}$).

### 4.2. From Calibrated Leak to Number of Zodis

If we decompose the total brightness distribution into that of the central star and the exozodi cloud, $I = I_\star + I_{zodi}$, it follows that:

$$
\begin{aligned}
L^{zodi}_{calculated} &\simeq \frac{\int\int I_{zodi}(x,y,\lambda)\cdot T(x,y,u,v,u_{xc},v_{xc},\lambda)^{null}dxdy}{\int\int I_\star(x,y,\lambda)\cdot T(x,y,u,v,u_{xc},v_{xc},\lambda)^{peak}dxdy} \\
&= \frac{\int\int I_{zodi}(x,y,\lambda)\cdot T(x,y,u,v,u_{xc},v_{xc},\lambda)^{null}dxdy}{F_\star}
\end{aligned}
\qquad (7)
$$

where the approximation holds when the stellar flux ($F_\star$) dominates over the exozodi flux, which is always the case here. The calculation is made monochromatically, at the effective wavelength (8.5 $\mu$m) of the $8 - 9\,\mu$m spectral bin used. The error introduced by this approximation is negligible, at the level of a fraction of a zodi, compared with typical 100s of zodi errors from the formal and external leak errors.

Thus, we also remove the stellar contribution from the measured calibrated leak:

$$
L^{zodi}_{measured} = (L_{calibrated} - L_\star) \qquad (8)
$$

where the leak due to the central star (Table 3 ) is calculated using an estimate of its angular diameter, following Serabyn (2000):

$$
L_\star \simeq \frac{\pi^2}{16}\cdot\left(\frac{B_p(m)\cdot\theta_\star(\mu rad)}{\lambda(\mu\text{m})}\right)^2 \qquad (9)
$$

where $B_p$ is the length of the projection of the baseline on the sky in the direction to the star and $\theta_\star$ is the stellar disk angular diameter (Table 1). We note that in the above expression we have neglected limb–darkening terms, unimportant for the spectral type and luminosity class of the stars in our sample.

Equation 8 illustrates an important advantage of the KIN measurements compared to spectro–photometric flux excess measurements which require accurate calibration of the stellar photosphere flux level: the excess leak ($L^{zodi}_{measured}$) estimate depends only on the stellar leak estimate ($L_\star$), which is easy to determine given the spectral types of the stars in our sample. Furthermore, the stars are generally small ($\sim$ 1 mas, Table 1) compared to the KIN resolution at 8.5 $\mu$m. Therefore the correction is small and the uncertainty in the stellar angular diameters affects very weakly the uncertainty on $L_\star$ and on $L^{zodi}_{measured}$.

For each target and for each individual observation, our procedure consists of the following steps:

1. Compute the measured excess leak $L^{zodi}_{measured}$. The uncertainty in the angular diameter of each of our target stars propagates into the uncertainty in the excess leak.

2. For a given exozodi disk orientation (inclination $i_{disk}$ and position angle $\mathrm{PA}_{disk}$), compute $L^{zodi}_{calculated}$, for a *Zodipic* model which has a dust density (number of zodis, $z$) such that the measured excess leak is matched: $L^{zodi}_{calculated} = L^{zodi}_{measured}$. Derive uncertainties in the required number of zodis from the formal and external errors. As noted earlier, this conversion to number of zodis must be made for each observation, because the instantaneous KIN fringe pattern that corresponds to each observation must be applied when using Equation 7.

3. Repeat for a range of disk inclinations and position angles which span the extremes in predicted leak; namely: face–on ($i_{disk} = 0°$) and edge–on ($i_{disk} = 90°$) with position angle parallel and perpendicular to the instantaneous direction of the long baseline fringes. In our averaging scheme, described below, we will take the variation resulting from the uncertainty in the exozodi cloud orientation as an additional source of error in the derived numbers of zodis. We note that (as can be seen in Table 4) in most cases the resulting uncertainty is significantly smaller than that due to the formal and external errors. As described in Section 5, for $\eta$ Crv and $\gamma$ Oph, imaging observations have constrained the inclination and position angle of the outer disk, and in those cases we also derive the number of zodis implied by our measurements by assuming that the exozodi disk has the same orientation as the outer disk.

Table 4 shows our results in terms of number of zodis ($z_{ij}$) for each observation and for each disk orientation. The notation refers to observation $i$ in "cluster" $j$. The notion of cluster must be introduced now in order to properly account for the external error in the averaging process described below. KIN observations of a given target begin with an adjustment to the static internal optical delay, and an on–sky alignment procedure which maximizes the mid–infrared flux seen by the nulling camera. We call a cluster a certain number of observations (typically $1-4$) of the same target in between such changes to the instrument setup. Clusters of observations of the same target may be made during the same night, or on distinct nights. Table 4 also shows the error in the number of zodis that result from the formal and external leak errors ($\sigma_{ij}^{formal}$ and $\sigma_{j}^{ext}$) as well as the error due to the uncertainty in the stellar leak ($\sigma^{\star}$). We note that although unphysical, negative zodis are allowed as a result of the error bars on the leak measurements. Finally, in order to describe the spatial extent of the exo–zodi region that the KIN is most sensitive to, the last column of Table 4 gives the half–light radius ($R_{\text{half-light}}$) of the azimuthally averaged exo–zodi brightness transmitted through the KIN fringe pattern. These radii may be compared, for example, to the radii of center of habitable zone given in Table 1.

## 4.3. Special Cases

### *4.3.1. Altair (α Aql)*

As mentioned above, the stellar diameters of our target stars are generally small, resulting in small errors in the stellar leak corrections. The exception is Altair ($\alpha$ Aql) which is large enough that uncertainties in its angular diameter can result in a significant uncertainty in the estimate of the excess leak. In order to estimate the stellar leak for this star, we use the image of its (elongated) photospheric disk made at the CHARA array (Monnier et al. 2007). The position angle of the KI baseline at the time of our observations ranged from 36° to 41°, approximately the same as the position angle of the major–axis of the CHARA image. Therefore, we use their measurement of the equatorial diameter ($3.6 \pm 0.016$ mas) in order to predict and correct for the stellar leak ($L_\star = 0.0172 \pm 0.00015$ and $0.0179 \pm 0.00015$, for our UT 2008 June 25 and 26 epochs respectively).

### *4.3.2. Binaries*

As can be seen in Table 1, five of our targets are known to have stellar companions (Kx Lib, 70 Oph, $\iota$ Peg, 61 Cyg A and 107 Psc). Therefore, we must evaluate their possible contribution to the measured leak, and correct for it if needed.

Two of those stars, Kx Lib and 70 Oph, have their companions at separations of 25 and 5 arcsec respectively, well outside the KIN field–of–view, and therefore have no impact on the measured leak.

For $\iota$ Peg, we have used the orbit of Boden et al. (1999) to determine the companion location at the epochs of our observations. We compute the leak due to the primary+secondary stellar system using the primary and secondary stellar angular diameters of Boden et al. (1999) and a primary/secondary flux ratio of 4.0 at 8.5 $\mu$m, also estimated using the stellar diameters and effective temperatures derived by Boden et al. (1999). This results in a stellar leak (both components) of $L_\star = 0.0152$ and 0.0153 for the 14 July 2008 UT = 14.24 and 14.95 epochs, respectively.

For 61 Cyg A and 107 Psc, we also use published orbits (from the Sixth Catalog of Orbits of Visual and Binary Stars, by W. I. Hartkopf and B. D. Mason[4]) to compute the companion locations at the epochs of the KIN observations. These companions are inside the KIN FOV. However, we have found no information from which the components flux ratio could be derived. Therefore, we perform no correction for these systems, and the results presented here in terms of zodi limits are to be considered upper–limits (i.e. any subtraction from the measured leak due to the companion would result in a smaller inferred exozodi level).

### 4.4. Averaging Scheme

After converting each leak measurement to a number of zodis we average the multiple observations available for each target in order to reduce the measurement errors. Consistent with the characterization of the external error described in Colavita et al. (2009, 2010), we average observations and clusters as follows:

1. Within each cluster: Compute the weighted mean of the number of zodis, $z_j = \sum(z_{ij} \cdot w_{ij})/\sum w_{ij}$, where the formal errors provide the weights, $w_{ij} = 1.0/\sigma^2_{ij,formal}$. The error in the weighted mean is taken to be the largest of the statistical error in the mean ($\sigma^a_{z_j} = 1.0/\sqrt{\sum w_{ij}}$) and the weighted average standard deviation in the data ($\sigma^b_{z_j} = \sqrt{\frac{\sum(z_{ij}-z_j)^2 \cdot w_{ij}}{\sum w_{ij}}}$). To this error we add the external error in quadrature: $\sigma_{z_j} = \sqrt{\sigma^2_{j,ext} + max\{\sigma^a_{z_j}, \sigma^b_{z_j}\}^2}$.

[4]http://ad.usno.navy.mil/wds/orb6.html

2. To combine the clusters: we again compute the weighted mean ($\overline{z} = \frac{\sum(z_j \cdot w_j)}{\sum w_j}$, with $w_j = 1.0/\sigma^2_{z_j}$), and the error in the weighted mean given by the statistical error ($\sigma_{\overline{z}} = 1.0/\sqrt{\sum w_j}$).

   We note that these first two steps imply that (a) the total error per cluster can never be smaller than the external error, and (b) when averaging multiple clusters, the total error can become smaller than the external error. The latter however generally provides only a small improvement, because our data sets for each target contain only 1 or 2 clusters.

3. The steps above are done for each exozodi orientation, the last step is to average the results over the 3 disk orientations computed ($i_{disk} = 0°$ and $i_{disk} = 90°$ with PA$_{disk}$ parallel or perpendicular to the instantaneous long baseline fringes). Thus, the final error contains a term (added in quadrature) computed as the rms of the number of zodis deduced for the 3 disk orientations ($\sigma_{[i_{disk}, \mathrm{pa}_{disk}]}$). As mentioned above, the "disk orientation error" is typically relatively small, tens of zodis, compared to that resulting from the formal and external error sources. The final number of exozodis for each target, including all data points and uncertainties is: $z = \sum_1^3(\overline{z})/3$, $\sigma_z = \sqrt{\sigma^2_{\overline{z}} + \sigma^2_{[i_{disk}, \mathrm{pa}_{disk}]}}$.

## 5. Results

Our final average results per target ($z$, $\sigma_z$) are shown in Table 5. The table also shows the detection significance in each case, as well as $3\sigma$ upper–limits, for ease of comparison with previous surveys that use similar metrics. These results are also illustrated in Figure 4.

### 5.1. $\eta$ Crv

Our only significant detection is $\eta$ Crv ($1250 \pm 260$ zodis). This object was previously known to have high levels of circumstellar dust, including observations with Spitzer/MIPS at $70\,\mu$m (Beichman et al. 2006b), Spitzer/IRS/ at $10 - 35\,\mu$m (Chen et al. 2006) and VLTI/MIDI at $10 - 13\,\mu$m (Smith et al. 2009). Also, as can be seen in Figure 2, the KIN spectrum across the N–band has adequate SNR to resolve the $10\,\mu$m silicate feature and can be used to infer dust properties. A detailed analysis of these data is left to future work.

The outer disk in this object has been directly imaged (Wyatt et al. 2005; Matthews et al. 2010). Thus, instead of taking the disk inclination and position angle as free parameters, we may assume that the exozodi disk has the same orientation. Using the parameters from the Matthews et al. (2010) Herschel observations ($i_{disk} = 50°$ with $\mathrm{PA}_{disk} = 103°$), we find $z = 1300 \pm 160$.

### 5.2. $\gamma$ Oph

This star was previously known to have excess starting at $15\mu$m, and growing much larger at longer wavelengths (Su et al. 2008). It is noteworthy that the infrared excess luminosity is several times larger for $\gamma$ Oph than for $\eta$ Crv, while the KIN makes a clear detection for $\eta$ Crv but only a marginal detection ($\sim 2.6\sigma$) for $\gamma$ Oph; implying perhaps that the dust distributions are very different in these two objects.

The Spitzer images spatially resolve the outer disk, and imply $i_{disk} = 50°$ and $\mathrm{PA}_{disk} = 55°$. Assuming the same orientation for the exozodi disk we derive $z = 200 \pm 60$.

### 5.3. $\alpha$ Aql (Altair)

This star was not previously know to have warm circumstellar dust. The KIN results indicate a marginal detection ($3.0\sigma$). The KIN measures an excess leak of $\sim 0.8\%$. This level of excess would have been undetectable by IRAS at $12\,\mu$m, and no $8\,\mu$m excess measurements have been made with Spitzer. Therefore, assuming the dust temperature is relatively warm, a KIN detection would not be inconsistent with the lack of excess seen in previous measurements.

### 5.4. Non–detections

Excluding our only clear detection ($\eta$ Crv) and the two possible detections ($\gamma$ Oph and $\alpha$ Aql), our sample contains 22 non–detections. Table 5 shows the upper limits on the number of zodis for each star that can be derived from our measurements. The average $3\sigma$ upper limit is 570 zodis. We note that the weighted RMS data scatter for the non–detections (150 zodis) is similar to the mean data uncertainty (160 zodis), which we take as an indication that the errors in the individual measurements have been fairly assigned.

## 6. Discussion and Conclusions

### 6.1. Comparison with Spitzer/IRS

Spitzer/IRS established exozodi limits based on fractional excess flux measurements ($F_{dust}/F_{\star}$) made in the $8.5 - 12\mu$m band (Beichman et al. 2006a; Lawler et al. 2009). Using a sample of 203 stars, the detection rate was only 1%. Based on their estimated error in the measurement of the fractional excess fluxes of $\sim 1\%$ ($1\sigma$ ), the sample of non–detection was used to place $3\sigma$ limits on the number of zodis in the range $600 - 2700$ zodis, with an average of 1100 zodis. Because the IRS short wavelength band is very similar to that of the KIN, it is of interest to compare the results from the two surveys.

First, we note that the KIN measurement can also be expressed in terms of a fractional flux excess measurement; which enables a direct comparison of expected performances (the much larger FOV of Spitzer/IRS is not expected to enter the comparison, because dust located outside the KIN FOV will be too cold to significantly contribute N–band flux). Assuming perfect stellar cancellation, the numerator in Equation 5 (i.e. at null) equals a fraction $f$ of the zodi flux $F_{dust}$, where $f$ is the instantaneous fraction of the zodi flux removed by the KIN fringe pattern at null. Likewise, since $F_\star >> F_{dust}$, the denominator (i.e. at peak) is essentially equal to the flux from the central star, therefore:

$$\frac{F_{dust}}{F_\star} \simeq \frac{L}{f} \tag{10}$$

and the error on this estimate of the fractional flux excess obtained from the KIN data is:

$$\sigma_{\left(\frac{F_{dust}}{F_\star}\right)} \simeq \frac{\sigma_L}{f} \tag{11}$$

The factor $f$ can be easily computed for each observation as the ratio of the exozodi flux transmitted through the KIN pattern at null to the total exozodi flux (for any value of $z$, say $z = 1$). On average, $f = 0.4$, and varies by less than 5% for our sample, including the exozodi cloud orientation effects (inclination and position angle, relative to the KIN fringe pattern). Therefore, in this conversion, the KIN measurement errors, $\sigma_L = 0.002$ to 0.004, are degraded by a factor of $1/f \simeq 2.5$, to become $\sigma_{\left(\frac{F_{dust}}{F_\star}\right)} = 0.005$ to to 0.0075. This is to be compared with the best–case IRS errors, $\sigma_{\left(\frac{F_{dust}}{F_\star}\right)} = 0.01$. It follows that one expects the current KIN implementation to provide up to ×2 tighter exozodi limits than Spitzer/IRS. We note however that the errors quoted for Spitzer/IRS do not include possible and potentially significant systematic errors arising from uncertainties in the absolute calibration of the stellar flux, to which the KIN is immune, as described earlier. In that sense, the above comparison represents a lower bound to the improvement that can be expected from the KIN.

Table 6 summarizes the detailed comparison of exozodi limits for the eight objects in common between the Spitzer/IRS and KIN surveys. It can be seen that on a target–by–target basis, either Spitzer/IRS or KIN can provide tighter limits, depending on the precise measurement errors in each case; but that we recover a common 2× improvement from KIN, as predicted.

### 6.2. Population Analysis

Having determined exozodi limits for each star in our sample, we now interpret those observations in terms of a statistically–defined exozodi disk model, in order to extract as much information as possible from the KIN measurements. Noting that the measured values in Table 5 tend to scatter around zero, and that the typical uncertainties are large compared to most of the measured values, and certainly large compared to the average of the measured values, we consider in this section the possibility that the true underlying exozodi values might be significantly smaller than the upper limits implied by the individual measurements.

For this analysis, we define the sub–sample of $N = 23$ stars which were not previously known to contain circumstellar dust (excludes $\eta$ Crv and $\gamma$ Oph). We assume that this sub–sample is representative of a single class of stars, with respect to the level of warm exozodi emission. In Appendix A we show that the calibrated net leaks are not correlated with any of the parameters in an exhaustive list describing the instrumental conditions. Furthermore, correlations are not expected among targets. Therefore, we assume that the zodi levels inferred for each star form an uncorrelated set, and that it is thus appropriate to use the 23 measurements to estimate the mean zodi level and its error for the class. We choose the bootstrap method to form a robust estimate of the mean and its error, independent of any assumptions on the underlying statistics (Efron & Tibshirani 1998); using with $10^6$ re–samples of the set of 23 measurements we find $\hat{z} = +2 \pm 50$ ($1\sigma$ error)[5].

Further interpretation of this result, in terms of a confidence level for the mean exo–zodi emission for the class being below a certain level, would require knowledge, which we do not have, of the underlying probability distribution for the measurement errors (i.e. whether or not they are approximately Gaussian) and of the underlying distribution of exo–zodi levels for the class of stars represented by our sample. If the measurement errors were Gaussian, the above result would imply that the mean exo–zodi level for the class is $< 50$ zodis ($1\sigma$ upper limit) with 84% confidence, or $< 150$ zodis ($3\sigma$ upper limit) with 99.8% confidence. Moreover, as detailed in Appendix B, we have also used Monte–Carlo simulations, and a wide range of assumed exo–zodi distribution, to show that these conclusions are largely insensitive to the shape and mean level of the true underlying exo–zodi distribution.

---

[5]This estimate of the error in the mean agrees well with one computed instead as the weighted RMS (170) divided by $\sqrt{\mathrm{EDOF}}$, resulting in $\sigma_{\hat{z}} = 58$, and where EDOF is the equivalent degrees of freedom which takes into consideration the data weights ($w_i$), EDOF $= (\sum w_i)^2)/\sum(w_i^2) = 9$.

### 6.3. Conclusions

Both our limits on individual stars, and the statistical result just discussed, are very positive for the future of direct imaging of exoplanets in the habitable zones around nearby stars, since they suggest that the exozodi levels might not be as high as was once feared. However, to be able to detect an exoEarth with any technique, the noise in a resolution element stemming from exozodi emission must be no more than a few Earth fluxes; and the limits presented here could still allow actual exozodi levels well above what can be tolerated. Measurements with sensitivity to lower exozodi levels are thus needed. The next step will likely be enabled by the Large Binocular Telescope Interferometer (LBTI), which will also use the nulling technique at mid–infrared wavelengths (Hinz 2009). Beyond that, a dedicated space mission, and multi–wavelength characterization, might be required.

## A. Correlation Analysis

In order to validate our noise model, we have examined the dependence of the calibrated net leaks (i.e. stellar contribution subtracted) against a set of 26 variables describing the instrumental and environmental conditions, and the astrophysical properties of the targets stars and associated calibrators. For this study, we have selected the (68) independent measurements corresponding to the sub–sample of 23 stars not previously known to contain circumstellar dust (excludes $\eta$ Crv and $\gamma$ Oph). The independent variables considered are: (1) day of year, (2) local time, (3) hour angle, (4) telescope azimuth, (5) telescope elevation, (6) internal optical delay difference, (7) right ascension, (8) declination, (9) stellar age, (10) stellar effective temperature[6], (11) detected N–band flux, (12) detected K–band

[6]To be sure, the measured leaks can be expected to correlate with stellar properties such as age or effective temperature; those parameters are tested in the event that they indirectly

flux (fringe tracker), (13) detected J–band flux (angle tracker), (14) angle tracker centroid RMS, (15) percent of leak data accepted by the quality gates, (16) precipitable water vapor (PWV), (17) wind speed, (18) wind direction, (19) air temperature, (20) relative humidity, (21) atmospheric pressure, (22) coherence time due to dry air (as in Colavita 2010), (23) coherence time due to water vapor (as in Colavita 2010), (24) calibrator diameter, (25) target–calibrator angular separation, and (27) target–calibrator flux ratio. The simultaneous PWV measurements are from the Caltech Submillimeter Observatory (CSO) tau–meter archive. All other atmospheric quantities are from the Canada–France–Hawaii Telescope (CFHT) weather archive.

Visual inspection of the correlation plots reveals that a linear relation would be sufficient to describe any candidate correlation. In order to obtain robust estimates of the statistical significance of any correlation candidate, we adopt a bootstrap approach. For each item, we perform a linear fit for each of $10^6$ bootstrap re–samples of the data. For each item, the mean of the $10^6$ fitted slopes is a good estimate of the slope; and the standard deviation is a good estimate of its uncertainty, independent of any assumptions on the underlying statistics (Efron & Tibshirani 1998). The ratio of the mean slope and error thus represents the significance of the slope being different from zero.

The result is that for most items (23 of the 26) the slope deviates from zero at a level less than $1\sigma$; and for all items the deviation is less than $2\sigma$. We therefore conclude that there are no statistically significant correlations in our data between the calibrated net leaks and any of the instrument/environment/astrophysical parameters that we have considered.

---

reveal a correlation with an instrument condition.

## B. Dependence of the mean exo–zodi confidence intervals on the underlying exozodi distribution

In order to determine the precise meaning of the error in the mean exo–zodi level calculated in Section 6.2, and to determine its dependence on the shape of the underlying distribution describing the number of stars with a certain exo-zodiacal level, we consider two illustrative examples: a uniform distribution ($f(z) = 1$ for $z$ in the interval $[0, 2\overline{z}]$, and 0 otherwise), and a half–exponential function ($f(z) = e^{-z/\overline{z}}$ for $z$ in the interval $[0, \infty]$), where $\overline{z}$ is the true mean exozodi level. We consider those distributions over a large range of $\overline{z}$ values, from 1 to 1000 zodis. The shapes of those two distributions (from completely flat to very centrally peaked) together with the large range of $\overline{z}$ values explored ensures our conclusions are robust with respect to the range of possible exozodi distributions.

Our procedure is as follows: (a) assume one of the exozodi distributions, uniform or half–exponential; (b) assume a value of the true mean zodi for the distribution ($\overline{z}$); (c) draw 23 random numbers ($z_i$) from the resulting $f(z)$ distribution; (d) simulate the measurement process by creating a simulated dataset by selecting 23 random numbers from normal distributions, of means = $\{z_i\}$ (from step (c) above) and standard deviations = $\{\sigma_z\}$ (the actual error on each exozodi datum); (e) compute the weighted mean and error in the mean for each set of simulated data; (f) repeat the simulation of the 23 measurements a large number of times ($10^6$), and count the number of times that the simulated mean differs from the true exozodi value ($\overline{z}$) by less than 1 or 2 times the error in the mean. The above procedure is repeated for $\overline{z}$ = 1 to 1000, and for both the uniform and half–exponential distributions.

By counting how many realizations of our simulated measurement process correspond to a true mean zodi value that lies within the range $\pm n\sigma$ from the mean, we generate a probability that this event could occur. The results are shown in Figure 5 where this probability is plotted for a range of true mean zodi values, and for both the uniform and an half–exponential distributions. These two distributions give essentially identical probabilities – 67% and 94% for 1 or $2\sigma$, respectively – for small values of the mean zodi level, and very similar results for larger values. This implies that the shape of the underlying distribution is not very important. We note that in Figure 5 the probability of the true mean value being in the indicated range falls off for large values of the true mean zodi. This is expected because in this case the distribution function is no longer being sampled densely enough over its whole range, so $N = 23$ samples are not quite sufficient to define the mean value. However this part of the curve is not important to our conclusions, because our interest is at smaller values.

The Keck Interferometer is funded by the National Aeronautics and Space Administration as part of its Exoplanet Exploration Program. The data presented herein were obtained at the W.M. Keck Observatory, which is operated as a scientific partnership among the California Institute of Technology, the University of California and the National Aeronautics and Space Administration. The Observatory was made possible by the generous financial support of the W.M. Keck Foundation. The authors wish to recognize and acknowledge the very significant cultural role and reverence that the summit of Mauna Kea has always had within the indigenous Hawaiian community. We are most fortunate to have the opportunity to conduct observations from this mountain. This work has made use of services produced by the NASA Exoplanet Science Institute at the California Institute of Technology. This research has made use of the Washington Double Star Catalog maintained at the U.S. Naval Observatory. Part of this work was carried out at the Jet Propulsion Laboratory, California Institute of Technology, under contract with NASA. The authors wish to acknowledge invaluable contributions from the KI team at Keck Observatory, the Jet Propulsion Lab and the NASA Exoplanet Science Institute. RMG acknowledges fruitful discussions with J. D. Monnier.

*Facilities:* Keck Interferometer.

Zuckerman, B. 2001, ARA&A, 39, 549

This manuscript was prepared with the AAS LaTeX macros v5.2.

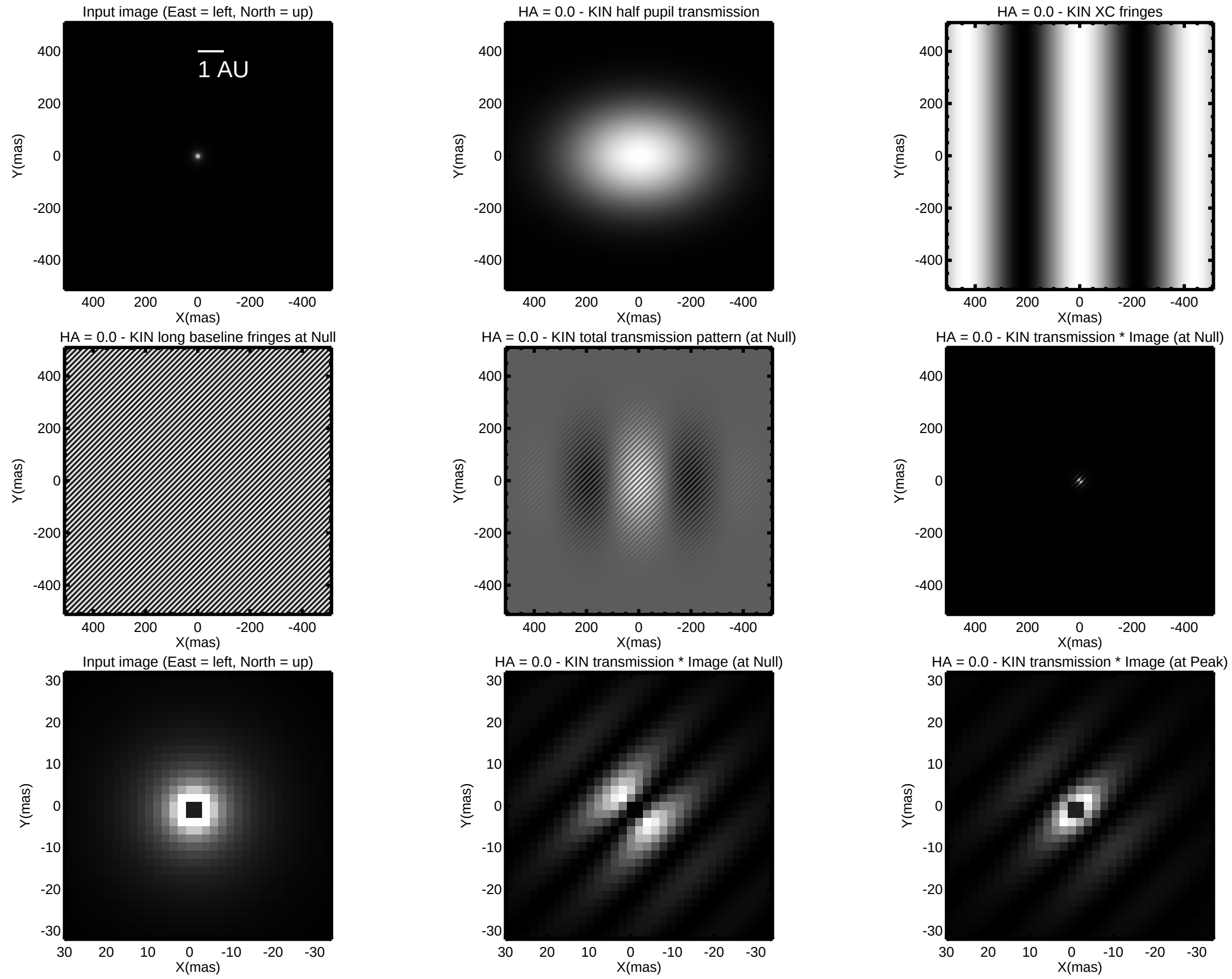

Fig. 1.— *Zodipic* image of a face–on Solar System analog at 10 pc, and KIN fringe patterns computed at 8.5 $\mu$m for H.A.=0 and dec=20°. In these images, the central star has been removed. Top–row: input image ($I$), half–aperture PSF ($T_{PSF}$), and short baseline fringes ($T_{XC}$). Middle–row: long baseline fringes ($T_{long}$), total transmission pattern (product of $T_{PSF} \times T_{XC} \times T_{long}$), and total transmitted brightness ($I \times T_{PSF} \times T_{XC} \times T_{long}$). Bottom–row: zoomed versions of the input image and transmitted brightness at null and peak.

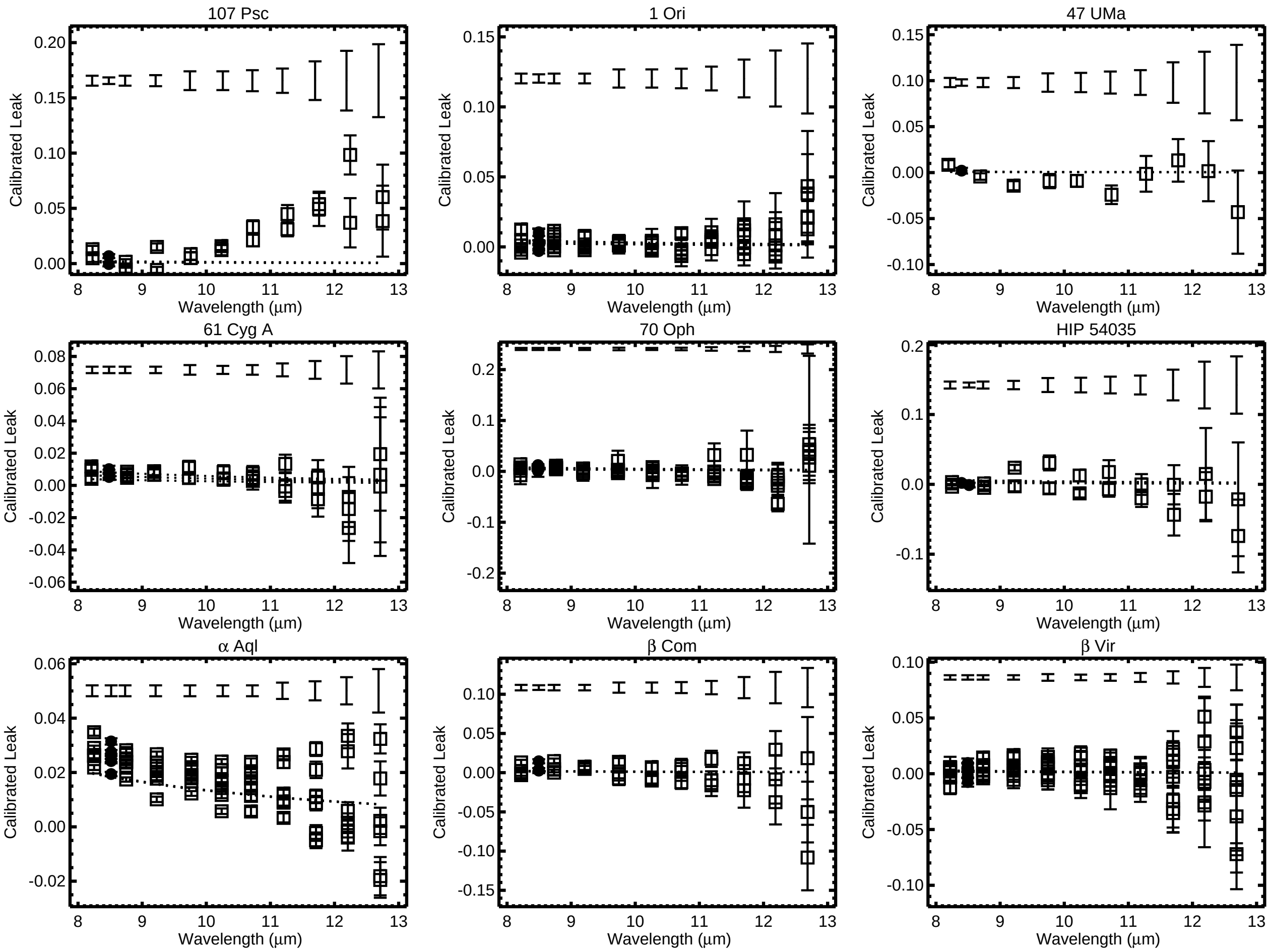

Fig. 2.— Calibrated leak data for all targets as a function of wavelength. Individual observations for each target, including formal errors, are shown as square symbols. The external errors for each target and at each wavelength are shown by the bars at the top of each plot. The leak level due to the central stars, and its uncertainty, are also shown (dotted lines). The degradation of the calibration quality at the red end of the bandpass, as discussed in the text, can be clearly seen. For the analysis presented in this paper we use only the $8-9\,\mu$m spectral bin (solid circles), which has the highest sensitivity for exozodi detection.

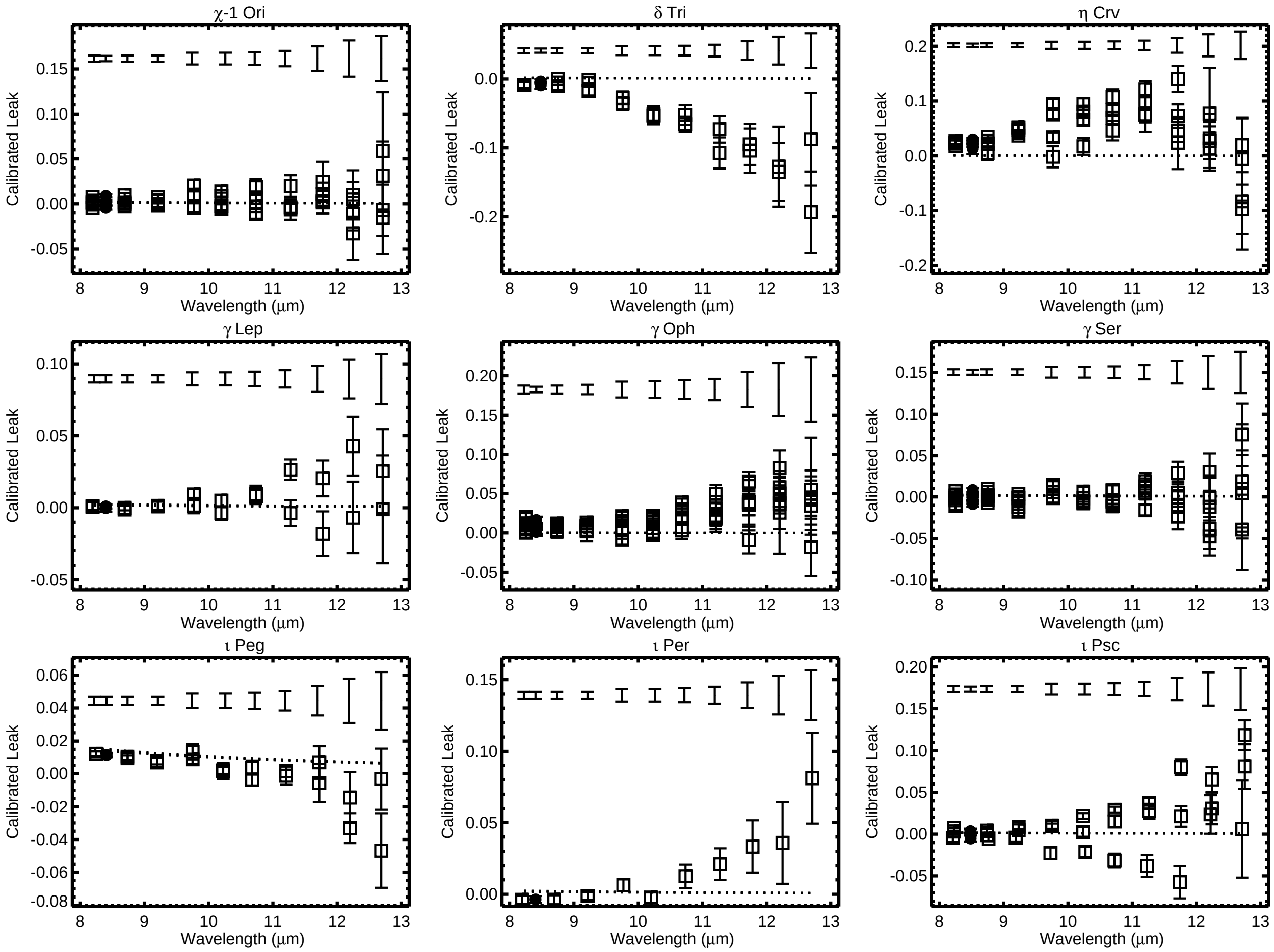

Fig. 2.— *continued*

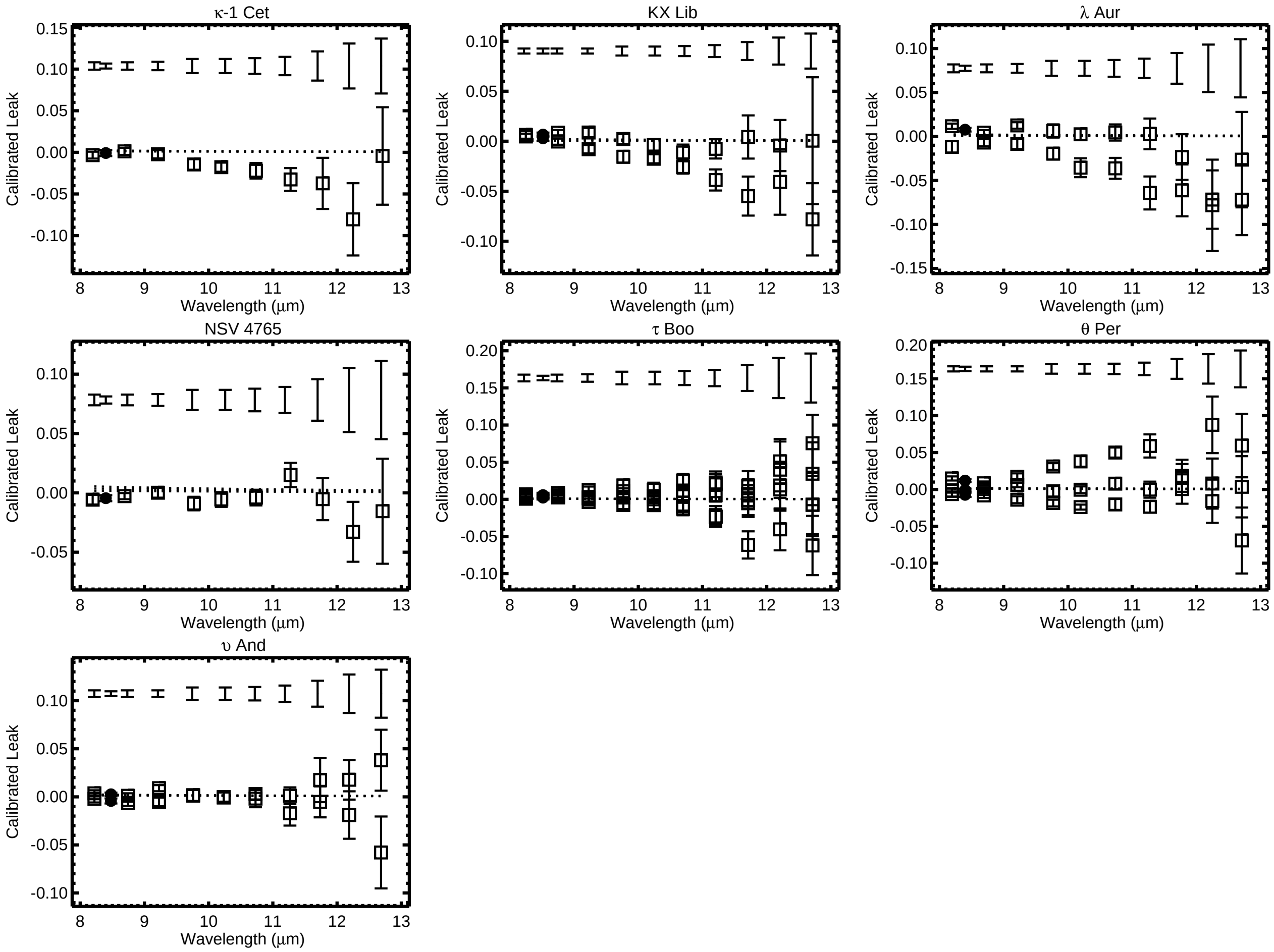

Fig. 2.— *continued*

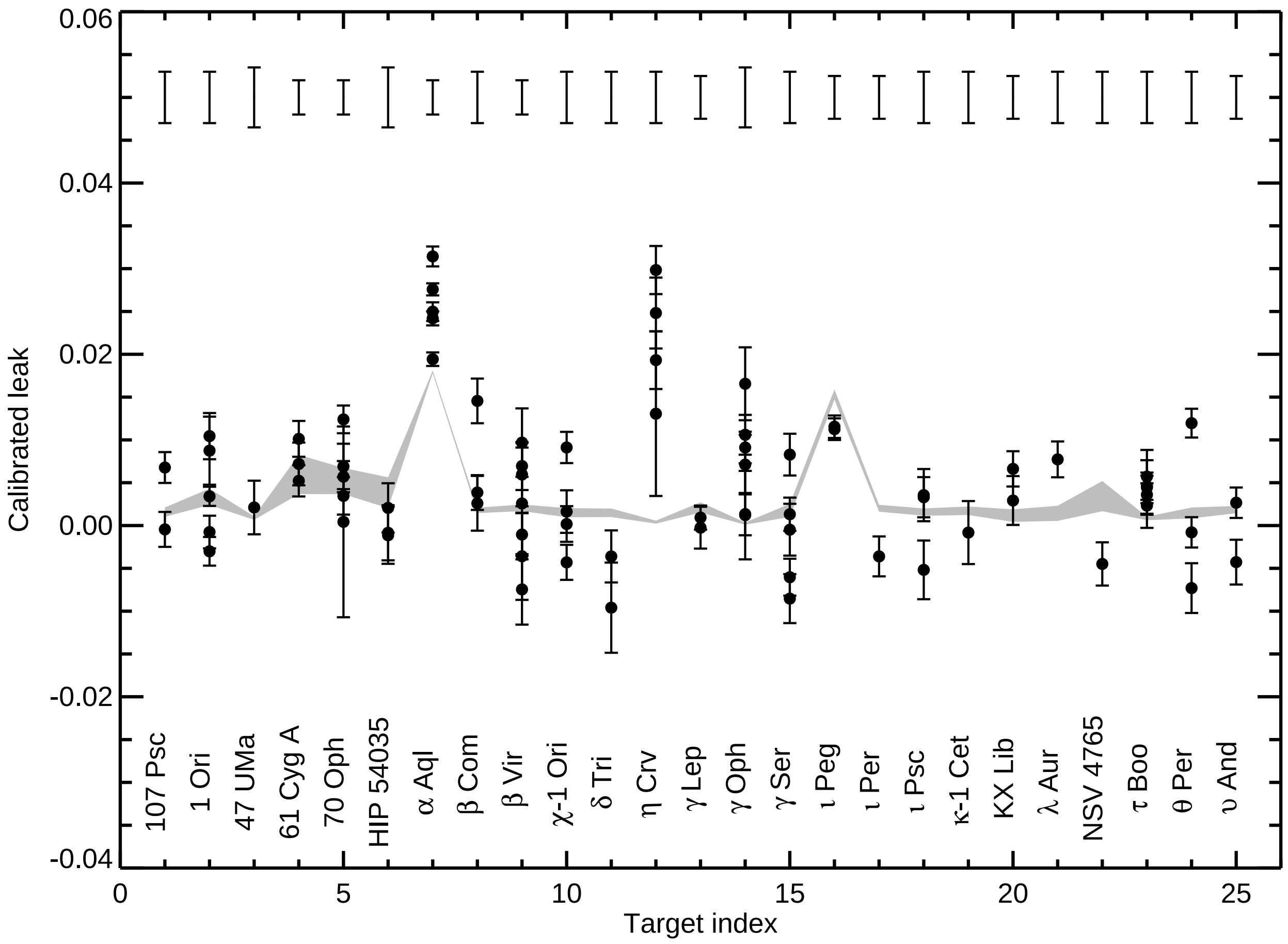

Fig. 3.— Calibrated data for the wideband $8-9\,\mu$m channel only. As in the previous figures, the external errors are shown as the bars at the top of the figure. The leak level due to the central stars, and its uncertainty, are also shown (grey band). We note that, as described in the text, at this stage in the analysis the multiple leak measurements for each target are not averaged, because they are allowed to vary with time as Earth rotation changes the baseline length and orientation.

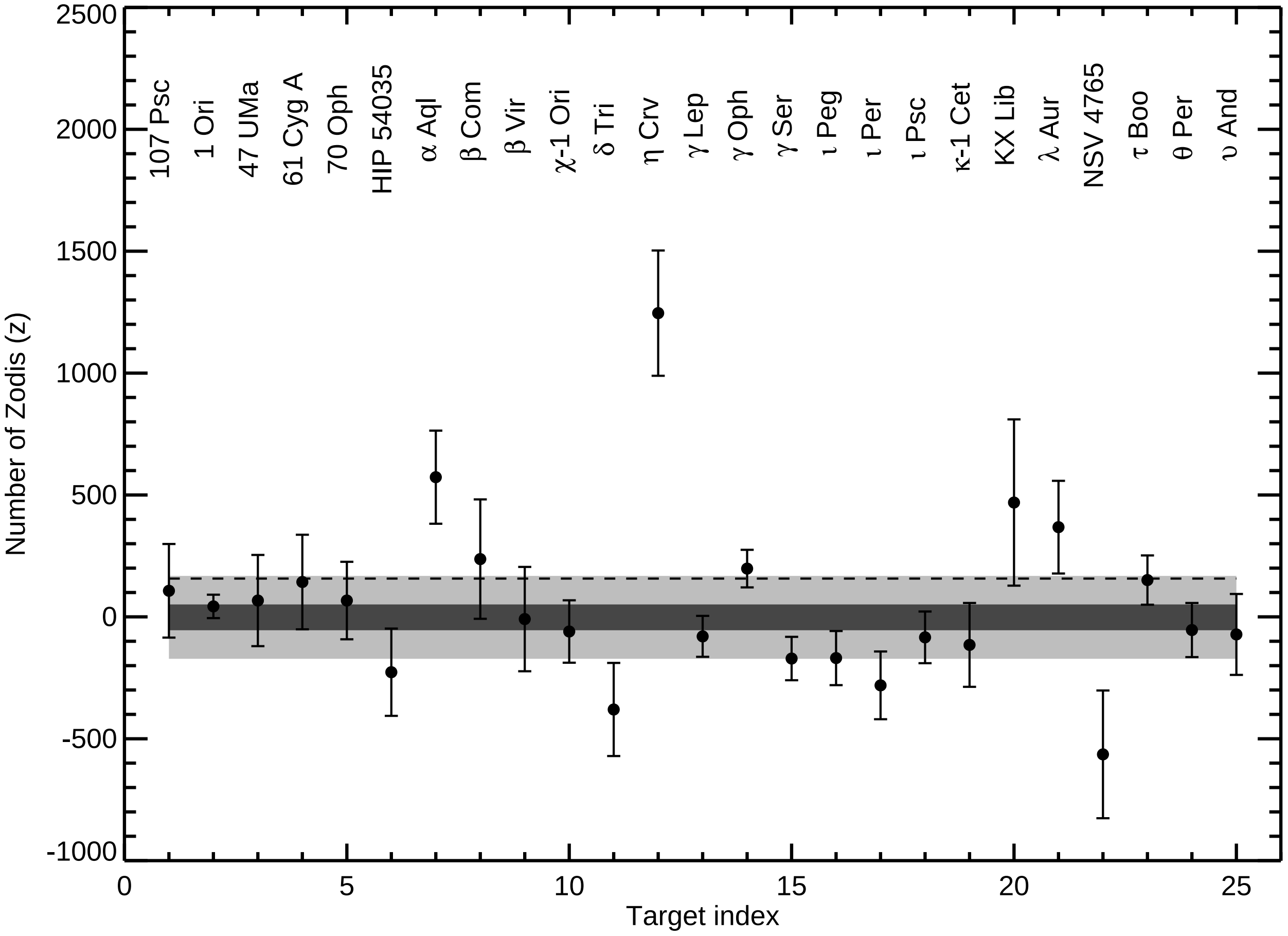

Fig. 4.— Average exozodi level per target. For the 22 clear non–detections, the weighted data scatter is 150 zodis, similar to typical measurement errors (mean 160 zodis), indicating that they have been accurately determined (see Section 5.4). For the sub–sample of 23 stars not previously known to have circumstellar dust (excludes $\eta$ Crv and $\gamma$ Oph) the mean is $\hat{z} = +2$, and the weighted scatter is $\sigma = 170$ zodis. The light grey band covers the range $\hat{z}\pm\sigma$. As discussed in Section 6.2, under the assumption that these 23 stars are representative a class from the point of view of the exo–zodi emission, and if the individual measurements are uncorrelated, we measure a mean and error in the mean $+2 \pm 50$ zodis, and this range is shown by the dark–grey band. The dashed line represents the $3\sigma$ upper–limit (150 zodis) inferred for the mean exo–zodi level for the class.

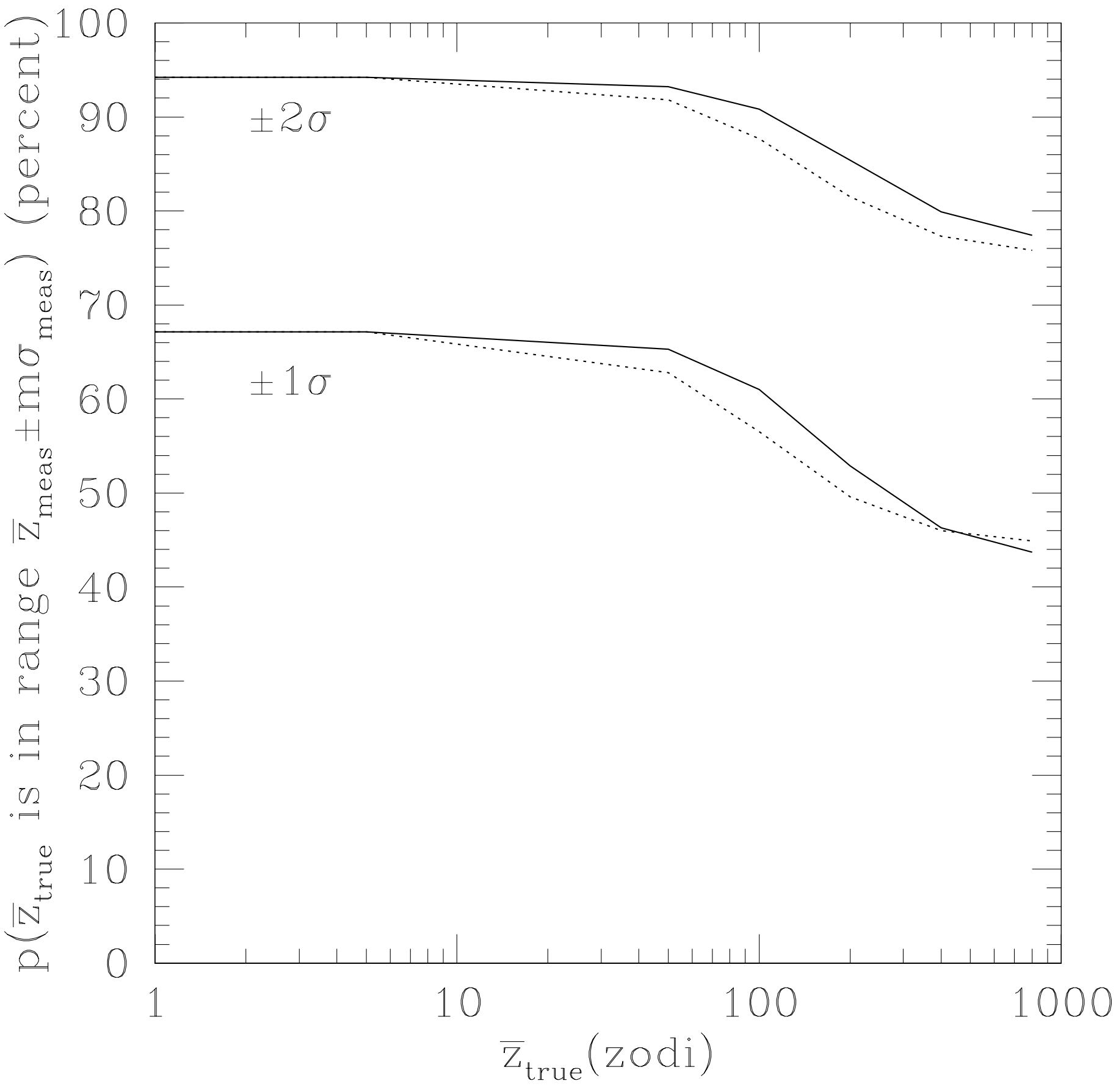

Fig. 5.— Probability that the true mean zodi level is in the range given by the simulated values of the mean plus or minus 1 (bottom curves) or 2 (top curves) times the error in the mean, assuming a uniform distribution of true zodi values (solid line) or a half–exponential distribution (dotted line).

Table 1. Target list.

| Name | | Spectral Type | RA | Dec | V | J | K | N (Jy) | d (pc) | R⋆ ($R_\odot$) | L⋆ ($L_\odot$) | T⋆ (K) | $\theta_\star$ (mas) | Age (Gyr) | $R_{\rm HZ}$ (AU) | Comments | R |
|---|---|---|---|---|---|---|---|---|---|---|---|---|---|---|---|---|---|
| 107 Psc | HIP7981 | K1V | 01:42:29.7 | +20:16:06.6 | 5.2 | 3.8 | 3.3 | 1.6 | 7.47 | 0.81 | 0.428 | 5180 | 1.1 ± 0.2 | 2.9 – 13.0 | 0.85 | Binary | 1 |
| 1 Ori | HIP22449 | F6V | 04:49:50.4 | +06:57:40.6 | 3.2 | 2.0 | 1.6 | 2.4 | 8.03 | 1.30 | 2.629 | 6450 | 1.5 ± 0.2 | 0.8 – 1.9 | 1.91 | ··· | 1 |
| 47 UMa | HIP53721 | G1V | 10:59:27.9 | +40:25:48.9 | 5.1 | 4.0 | 3.7 | 1.3 | 14.1 | 1.22 | 1.555 | 5860 | 0.8 ± 0.1 | 4.2 – 5.9 | 1.54 | 3 RV planets | 1 |
| 61 Cyg a | HIP104214 | K5V | 21:06:53.9 | +38:44:57.9 | 5.2 | 3.1 | 2.2 | 5.5 | 3.48 | 0.74 | 0.167 | 4300 | 2.0 ± 0.4 | ··· | 0.57 | Binary | · |
| 70 Oph | HIP88601 | K0V | 18:05:27.3 | +02:30:00.3 | 4.0 | 2.3 | 1.8 | 7.0 | 5.09 | 0.99 | 0.615 | 5140 | 2.2 ± 0.3 | 1.1 – 1.9 | 1.03 | Binary | 3 |
| HD95735 | HIP54035 | M2V | 11:03:20.2 | +35:58:11.5 | 7.5 | 4.2 | 3.2 | 1.3 | 2.55 | 0.34 | 0.020 | 3730 | 1.7 ± 0.4 | ··· | 0.20 | ··· | · |
| $\alpha$ Aql | HIP97649 | A7V | 19:50:47.0 | +08:52:05.9 | 0.8 | 0.3 | 0.1 | 31.0 | 5.14 | 1.79 | 9.845 | 7650 | 3.1 ± 0.5 | 0.7 | 3.33 | ··· | 4 |
| $\beta$ Com | HIP64394 | G0V | 13:11:52.3 | +27:52:41.4 | 4.3 | 3.2 | 2.9 | 2.4 | 9.15 | 1.09 | 1.341 | 5960 | 1.1 ± 0.1 | 0.5 – 3.0 | 1.42 | ··· | 1 |
| $\beta$ Vir | HIP57757 | F9V | 11:50:41.7 | +01:45:52.9 | 3.6 | 2.6 | 2.3 | 4.3 | 10.90 | 1.68 | 3.443 | 6080 | 1.5 ± 0.1 | 2.3 – 3.4 | 2.26 | ··· | 1 |
| $\chi$1 Ori | HIP27913 | G0V | 05:54:22.9 | +20:16:34.2 | 4.4 | 3.4 | 3.0 | 2.4 | 8.66 | 0.98 | 1.05 | 5930 | 1.1 ± 0.2 | 1.4 – 5.5 | 1.26 | ··· | 1 |
| $\delta$ Tri | HIP10644 | G0.5V | 02:17:03.2 | +34:13:27.2 | 4.9 | 3.5 | 3.1 | 1.8 | 10.85 | 1.02 | 1.09 | 5860 | 1.0 ± 0.2 | 1.9 | 1.29 | ··· | 2 |
| $\eta$ Crv | HIP61174 | F2V | 12:32:04.2 | -16:11:45.6 | 4.3 | 3.6 | 3.4 | 1.8 | 18.21 | 1.53 | 4.679 | 6870 | 0.8 ± 0.2 | 0.95 | 2.46 | High dust | 5 |
| $\gamma$ Lep A | HIP27072 | F6V | 05:44:27.7 | -22:26:54.1 | 3.6 | 2.7 | 2.4 | 3.8 | 8.97 | 1.22 | 2.259 | 6410 | 1.3 ± 0.2 | 1.3 – 2.5 | 1.78 | ··· | 6 |
| $\gamma$ Oph | HIP87108 | A0V | 17:47:53.5 | +02:42:26.2 | 3.7 | 3.6 | 3.6 | 1.2 | 29.05 | 1.92 | 21.937 | 9030 | 0.5 ± 0.1 | 0.2 | 4.32 | High dust | 7 |
| $\gamma$ Ser | HIP78072 | F6IV | 15:56:27.1 | +15:39:41.8 | 3.8 | 3.1 | 2.7 | 2.3 | 11.12 | 1.37 | 2.741 | 6370 | 1.2 ± 0.3 | 3.4 – 3.9 | 1.97 | ··· | 1 |
| $\iota$ Peg | HIP109176 | F5V | 22:07:00.6 | +25:20:42.4 | 3.8 | 2.9 | 2.6 | 2.8 | 11.76 | 1.42 | 3.319 | 6540 | 1.1 ± 0.2 | 1.97 | 2.13 | Binary | 5 |
| $\iota$ Per | HIP14632 | G0V | 03:09:04.0 | +49:36:47.7 | 4.0 | 3.1 | 2.7 | 2.9 | 10.53 | 1.42 | 2.176 | 5890 | 1.2 ± 0.1 | 3.0 – 3.3 | 1.82 | ··· | 1 |
| $\iota$ Psc | HIP116771 | F7V | 23:39:57.0 | +05:37:34.6 | 4.1 | 3.3 | 2.9 | 2.3 | 13.79 | 1.57 | 3.324 | 6240 | 1.0 ± 0.1 | 3.5 – 4.0 | 2.19 | ··· | 1 |
| $\kappa$1 Cet | HIP15457 | G5V | 03:19:21.6 | +03:22:12.7 | 4.8 | 3.4 | 2.9 | 1.6 | 9.16 | 0.97 | 0.832 | 5620 | 1.1 ± 0.2 | 0.6 – 4.1 | 1.66 | ··· | 1 |
| kx Lib | HIP73184 | K4V | 14:57:27.9 | -21:24:55.7 | 5.7 | 3.7 | 3.0 | 3.4 | 5.91 | 0.82 | 0.263 | 4570 | 1.4 ± 0.5 | 0.2 – 11.6 | 0.70 | Binary | 1 |
| $\lambda$ Aur | HIP24813 | G1.5IV | 05:19:08.4 | +40:05:56.5 | 4.7 | 3.4 | 3.0 | 1.9 | 12.65 | 1.30 | 1.726 | 5820 | 1.0 ± 0.3 | 3.3 – 4.7 | 1.63 | ··· | 1 |
| NSV4765 | HIP49908 | K5V | 10:11:22.1 | +49:27:15.2 | 6.6 | 3.9 | 2.9 | 1.9 | 4.87 | 0.79 | 0.133 | 3920 | 1.6 ± 0.4 | 4.7 | 0.52 | ··· | 8 |
| $\tau$ Boo | HIP67275 | F6IV | 13:47:15.7 | +17:27:24.8 | 4.5 | 3.6 | 3.5 | 1.6 | 15.60 | 1.43 | 2.99 | 6370 | 0.8 ± 0.1 | 0.7 – 1.7 | 2.05 | 1 RV planet | 1 |

Table 1—Continued

| Name | | Spectral Type | RA | Dec | V | J | K | N (Jy) | d (pc) | $R_\star$ ($R_\odot$) | $L_\star$ ($L_\odot$) | $T_\star$ (K) | $\theta_\star$ (mas) | Age (Gyr) | $R_{HZ}$ (AU) | Comments | Age Referen |
|---|---|---|---|---|---|---|---|---|---|---|---|---|---|---|---|---|---|
| $\theta$ Per | HIP12777 | F7V | 02:44:11.9 | +49:13:42.4 | 4.1 | 3.0 | 2.7 | 2.0 | 11.23 | 1.24 | 2.188 | 6320 | $1.0 \pm 0.2$ | $0.5 - 2.0$ | 1.77 | ··· | 1 |
| $\upsilon$ And | HIP7513 | F8V | 01:36:47.8 | +41:24:19.6 | 4.1 | 3.2 | 2.9 | 2.6 | 13.47 | 1.62 | 3.301 | 6120 | $1.1 \pm 0.1$ | $2.3 - 3.0$ | 2.20 | 3 RV planets | 1 |

Note. — $R_{HZ}$ is the radius of the center of the habitable zone. Stellar data from NStED. Diameters from NExScI/fBol (validated with surface brightness relations). Age referen 1: Valenti & Fischer (2005), 2: Wright et al. (2004), 3: Mamajek & Hillenbrand (2008), 4: Rieke et al. (2005), 5: Vizier online data catalog V/89 (Marsakov & Shevelev 1995) Holmberg et al. (2009), 7: Rhee et al. (2007), 8: Bryden et al. (2006)

Table 2. Calibrator Stars.

| Name | Spectral Type | $\theta_\star(K)$ (mas) | error (mas) | $\theta_\star(N)$ (mas) | error (mas) | Calibrator for |
|---|---|---|---|---|---|---|
| HD1635 | K3III | 1.593 | 0.09 | 1.619 | 0.10 | $\iota$ Psc |
| HD7106 | K0III | 1.871 | 0.10 | 1.901 | 0.11 | $\delta$ Tri |
| HD7147 | K4III | 1.273 | 0.13 | 1.293 | 0.14 | $\iota$ Psc |
| HD12594 | K4III | 1.578 | 0.15 | 1.603 | 0.15 | 107 Psc |
| HD13363 | K4III | 1.542 | 0.18 | 1.567 | 0.18 | 107 Psc |
| HD14770 | G8III | 1.226 | 0.21 | 1.245 | 0.21 | $\theta$ Per |
| HD15779 | G3III | 1.569 | 0.30 | 1.594 | 0.31 | $\kappa$1 Cet |
| HD16028 | K4III | 1.601 | 0.10 | 1.627 | 0.11 | $\upsilon$ And |
| HD16160 | K3V | 1.138 | 0.31 | 1.156 | 0.31 | $\kappa$1 Cet |
| HD18339 | K3III | 1.586 | 0.14 | 1.611 | 0.14 | $\theta$ Per, $\iota$ Per |
| HD23413 | K4III | 1.875 | 0.18 | 1.905 | 0.18 | 1 Ori |
| HD29317 | K0III | 1.686 | 0.19 | 1.713 | 0.19 | $\theta$ Per, $\iota$ Per |
| HD30557 | G9III | 1.116 | 0.16 | 1.134 | 0.17 | $\lambda$ Aur |
| HD34559 | G8III | 1.091 | 0.33 | 1.108 | 0.34 | $\chi$1 Ori |
| HD36780 | K5III | 2.017 | 0.12 | 2.049 | 0.12 | 1 Ori |
| HD36923 | M0III | 1.462 | 0.13 | 1.486 | 0.13 | $\gamma$ Lep |
| HD42341 | K2III | 1.141 | 0.10 | 1.159 | 0.10 | $\gamma$ Lep |
| HD42398 | K0III | 1.027 | 0.20 | 1.043 | 0.20 | $\chi$1 Ori |
| HD43993 | K1III | 1.484 | 0.10 | 1.507 | 0.10 | $\gamma$ Lep |
| HD45433 | K5III | 1.818 | 0.22 | 1.847 | 0.22 | 1 Ori |
| HD46374 | K2III | 1.363 | 0.19 | 1.385 | 0.19 | $\chi$1 Ori |
| HD46709 | K4III | 1.700 | 0.09 | 1.727 | 0.09 | $\chi$1 Ori |
| HD47070 | K5III | 1.255 | 0.23 | 1.275 | 0.23 | $\lambda$ Aur |
| HD93859 | K2III | 0.950 | 0.22 | 0.965 | 0.22 | NSV4765 |
| HD94669 | K2III | 1.098 | 0.15 | 1.115 | 0.15 | 47 Uma, HD95735 |
| HD95345 | K1III | 2.077 | 0.15 | 2.111 | 0.16 | $\beta$ Vir |
| HD95849 | K3III | 0.614 | 0.70 | 0.623 | 0.71 | $\eta$ Crv |
| HD99967 | K2III | 1.124 | 0.14 | 1.142 | 0.14 | 47 Uma, HD95735, NSV4765 |
| HD100343 | K4III | 1.432 | 0.15 | 1.455 | 0.15 | $\eta$ Crv |
| HD102159 | M4III | 4.232 | 0.09 | 4.299 | 0.09 | HD95735 |
| HD103500 | M3III | 2.526 | 0.18 | 2.566 | 0.18 | HD95735 |

Table 2—Continued

| Name | Spectral Type | $\theta_\star(K)$ (mas) | error (mas) | $\theta_\star(N)$ (mas) | error (mas) | Calibrator for |
|---|---|---|---|---|---|---|
| HD104979 | G8III | 1.981 | 0.09 | 2.012 | 0.09 | $\beta$ Vir |
| HD107325 | K2III | 1.088 | 0.25 | 1.105 | 0.26 | $\beta$ Com |
| HD107328 | K1III | 1.904 | 0.14 | 1.934 | 0.14 | $\beta$ Vir |
| HD107418 | K0III | 1.098 | 0.26 | 1.116 | 0.27 | $\eta$ Crv |
| HD109317 | K0III | 1.062 | 0.06 | 1.079 | 0.06 | $\beta$ Com |
| HD117818 | K0III | 0.931 | 0.39 | 0.946 | 0.40 | $\eta$ Crv |
| HD118840 | M3III | 1.241 | 0.12 | 1.261 | 0.12 | $\tau$ Boo |
| HD119126 | G9III | 1.005 | 0.16 | 1.021 | 0.16 | $\beta$ Com |
| HD119584 | K4III | 1.595 | 0.10 | 1.620 | 0.10 | $\tau$ Boo |
| HD124206 | K2III | 1.958 | 0.23 | 1.989 | 0.24 | kx Lib |
| HD129972 | K0III | 1.526 | 0.23 | 1.551 | 0.24 | $\gamma$ Ser |
| HD144889 | K4III | 1.266 | 0.11 | 1.286 | 0.11 | $\tau$ Boo |
| HD147547 | A9III | 0.907 | 0.13 | 0.921 | 0.13 | $\gamma$ Ser |
| HD151217 | K5III | 3.187 | 0.22 | 3.238 | 0.23 | 70 Oph |
| HD152601 | K2III | 0.998 | 0.23 | 1.014 | 0.23 | $\gamma$ Oph |
| HD166460 | K2III | 1.062 | 0.14 | 1.079 | 0.14 | $\gamma$ Oph |
| HD170474 | K0III | 0.886 | 0.15 | 0.900 | 0.16 | $\gamma$ Oph |
| HD171391 | G8III | 1.076 | 0.14 | 1.093 | 0.15 | $\gamma$ Oph |
| HD176678 | K1III | 2.867 | 0.18 | 2.913 | 0.18 | 70 Oph |
| HD184406 | K3III | 3.001 | 0.07 | 3.049 | 0.07 | 70 Oph, $\alpha$ Aql |
| HD198134 | K3III | 2.053 | 0.25 | 2.085 | 0.25 | $\iota$ Peg |
| HD199169 | K4III | 2.715 | 0.17 | 2.759 | 0.17 | $\iota$ Peg |
| HD205512 | K1III | 2.028 | 0.24 | 2.060 | 0.24 | 61 Cyg a |
| HD213119 | K5III | 2.332 | 0.08 | 2.369 | 0.08 | $\iota$ Psc |
| HD217459 | K4III | 1.326 | 0.10 | 1.347 | 0.10 | $\iota$ Psc |

Note. — Calibrator uniform disk angular diameters ($\theta_\star$, at K and N–band) are from Appendix A of Colavita et al. (2009).

Table 3. Log of Observations and Calibrated Leak Data (wideband channel $8 - 9\,\mu$m).

| Name | Date | HA | u (m) | v (m) | $L_{calibrated}$ | $\sigma^{formal}_{L_{calibrated}}$ | $\sigma^{ext.}_{L_{calibrated}}$ | $L_\star$ |
|---|---|---|---|---|---|---|---|---|
| 107 Psc | UT 2008 Oct 13 | 2.54 | 26.5162 | 76.8456 | 0.006772 | 0.001803 | 0.003000 | 0.001605 ± 0.000554 |
| 107 Psc | UT 2008 Oct 13 | 3.27 | 16.6299 | 78.2699 | -0.000451 | 0.002041 | 0.003000 | 0.001555 ± 0.000537 |
| 1 Ori | UT 2008 Feb 17 | 0.41 | 48.9782 | 69.3127 | -0.003011 | 0.001677 | 0.003000 | 0.003428 ± 0.000979 |
| 1 Ori | UT 2008 Feb 17 | 1.10 | 43.0509 | 72.2259 | 0.003409 | 0.001118 | 0.003000 | 0.003365 ± 0.000961 |
| 1 Ori | UT 2008 Feb 17 | 1.83 | 35.3784 | 74.8085 | -0.000761 | 0.001905 | 0.003000 | 0.003259 ± 0.000931 |
| 1 Ori | UT 2008 Feb 18 | 0.10 | 51.1001 | 67.9093 | 0.010442 | 0.002699 | 0.003000 | 0.003438 ± 0.000982 |
| 1 Ori | UT 2008 Feb 18 | 1.22 | 41.8864 | 72.6852 | 0.008744 | 0.003977 | 0.003000 | 0.003349 ± 0.000957 |
| 47 UMa | UT 2009 Jan 10 | -1.08 | 56.0337 | 53.2305 | 0.002110 | 0.003131 | 0.003500 | 0.000846 ± 0.000201 |
| 61 Cyg A | UT 2008 Aug 17 | 0.46 | 48.5433 | 67.6194 | 0.010119 | 0.002093 | 0.002000 | 0.005730 ± 0.002202 |
| 61 Cyg A | UT 2008 Aug 17 | 1.12 | 42.8783 | 72.5475 | 0.005219 | 0.001828 | 0.002000 | 0.005873 ± 0.002257 |
| 61 Cyg A | UT 2008 Aug 17 | 1.70 | 36.7988 | 76.3515 | 0.007192 | 0.002493 | 0.002000 | 0.005941 ± 0.002283 |
| 70 Oph | UT 2008 Aug 17 | 1.27 | 41.3929 | 65.0748 | 0.012394 | 0.001618 | 0.002000 | 0.005621 ± 0.001658 |
| 70 Oph | UT 2008 Aug 17 | 1.86 | 34.9364 | 65.3338 | 0.005707 | 0.001824 | 0.002000 | 0.005187 ± 0.001530 |
| 70 Oph | UT 2008 Aug 18 | 0.35 | 49.3793 | 64.5963 | 0.003461 | 0.002184 | 0.002000 | 0.006247 ± 0.001842 |
| 70 Oph | UT 2008 Aug 18 | 1.03 | 43.6993 | 64.9598 | 0.000431 | 0.011140 | 0.002000 | 0.005792 ± 0.001708 |
| 70 Oph | UT 2008 Aug 18 | 1.89 | 34.6930 | 65.3421 | 0.006900 | 0.002653 | 0.002000 | 0.005172 ± 0.001525 |
| HIP 54035 | UT 2008 Apr 14 | 2.75 | 23.7420 | 81.4917 | -0.000848 | 0.003220 | 0.003500 | 0.003984 ± 0.001872 |
| HIP 54035 | UT 2008 Apr 14 | 3.45 | 14.0496 | 83.5171 | -0.001139 | 0.003330 | 0.003500 | 0.003966 ± 0.001864 |
| HIP 54035 | UT 2009 Jan 10 | -0.23 | 53.0053 | 62.9383 | 0.002060 | 0.002876 | 0.003500 | 0.003744 ± 0.001759 |
| $\alpha$ Aql | UT 2008 May 25 | -1.22 | 56.2663 | 63.5521 | 0.027578 | 0.000705 | 0.002000 | 0.018738 ± 0.000167 |
| $\alpha$ Aql | UT 2008 May 25 | -0.42 | 53.9181 | 65.3231 | 0.024192 | 0.000818 | 0.002000 | 0.018658 ± 0.000166 |
| $\alpha$ Aql | UT 2008 May 26 | -0.93 | 55.6966 | 64.1969 | 0.024979 | 0.001091 | 0.002000 | 0.018786 ± 0.000167 |
| $\alpha$ Aql | UT 2008 May 26 | -0.25 | 53.0955 | 65.6963 | 0.031427 | 0.001162 | 0.002000 | 0.018557 ± 0.000165 |
| $\alpha$ Aql | UT 2008 May 26 | 0.35 | 49.4061 | 66.9415 | 0.019427 | 0.000794 | 0.002000 | 0.018003 ± 0.000160 |
| $\beta$ Com | UT 2008 Feb 16 | 0.20 | 50.4131 | 68.0667 | 0.014552 | 0.002608 | 0.003000 | 0.001774 ± 0.000320 |
| $\beta$ Com | UT 2008 Feb 16 | 1.08 | 43.2738 | 73.1044 | 0.002594 | 0.003199 | 0.003000 | 0.001784 ± 0.000322 |
| $\beta$ Com | UT 2008 Feb 16 | 1.87 | 34.8345 | 76.9074 | 0.003855 | 0.002029 | 0.003000 | 0.001762 ± 0.000318 |
| $\beta$ Vir | UT 2008 Feb 17 | 1.66 | 37.2863 | 64.7000 | -0.001053 | 0.002496 | 0.002000 | 0.002551 ± 0.000473 |
| $\beta$ Vir | UT 2008 Feb 17 | 2.33 | 29.2457 | 64.8762 | 0.006946 | 0.002802 | 0.002000 | 0.002317 ± 0.000430 |
| $\beta$ Vir | UT 2008 Feb 17 | 3.04 | 19.8380 | 65.0126 | 0.009681 | 0.004003 | 0.002000 | 0.002114 ± 0.000392 |
| $\beta$ Vir | UT 2008 Feb 18 | 1.55 | 38.4409 | 64.6688 | -0.003591 | 0.005091 | 0.002000 | 0.002590 ± 0.000480 |

Table 3—Continued

| Name | Date | HA | u (m) | v (m) | $L_{calibrated}$ | $\sigma^{formal}_{L_{calibrated}}$ | $\sigma^{ext.}_{L_{calibrated}}$ | $L_{\star}$ |
|---|---|---|---|---|---|---|---|---|
| $\beta$ Vir | UT 2008 Feb 18 | 2.56 | 26.3314 | 64.9254 | -0.007462 | 0.004110 | 0.002000 | 0.002246 ± 0.000416 |
| $\beta$ Vir | UT 2008 Feb 18 | 2.69 | 24.5754 | 64.9519 | 0.002577 | 0.006531 | 0.002000 | 0.002207 ± 0.000409 |
| $\beta$ Vir | UT 2008 Feb 18 | 3.25 | 16.8910 | 65.0429 | 0.005953 | 0.003708 | 0.002000 | 0.002066 ± 0.000383 |
| $\chi$-1 Ori | UT 2009 Jan 10 | 1.15 | 42.5623 | 72.4220 | 0.009114 | 0.001823 | 0.003000 | 0.001621 ± 0.000576 |
| $\chi$-1 Ori | UT 2009 Jan 10 | 1.90 | 34.5281 | 75.0401 | 0.000176 | 0.002080 | 0.003000 | 0.001568 ± 0.000557 |
| $\chi$-1 Ori | UT 2009 Jan 13 | 2.35 | 29.0718 | 76.3297 | 0.001628 | 0.002485 | 0.003000 | 0.001533 ± 0.000544 |
| $\chi$-1 Ori | UT 2009 Jan 13 | 3.35 | 15.5151 | 78.3664 | -0.004302 | 0.002044 | 0.003000 | 0.001466 ± 0.000521 |
| $\delta$ Tri | UT 2008 Nov 13 | 0.44 | 48.7305 | 68.5848 | -0.009594 | 0.005263 | 0.003000 | 0.001449 ± 0.000488 |
| $\delta$ Tri | UT 2008 Nov 13 | 1.95 | 33.9310 | 77.9062 | -0.003606 | 0.003037 | 0.003000 | 0.001478 ± 0.000498 |
| $\eta$ Crv | UT 2008 Apr 17 | -1.93 | 56.3123 | 62.3397 | 0.024819 | 0.004139 | 0.003000 | 0.000976 ± 0.000423 |
| $\eta$ Crv | UT 2008 May 24 | 0.71 | 46.6175 | 51.9715 | 0.029837 | 0.002808 | 0.003000 | 0.000674 ± 0.000292 |
| $\eta$ Crv | UT 2008 May 24 | 1.91 | 34.3913 | 48.3670 | 0.019313 | 0.003375 | 0.003000 | 0.000487 ± 0.000211 |
| $\eta$ Crv | UT 2008 May 24 | 2.61 | 25.6853 | 46.8307 | 0.013058 | 0.009607 | 0.003000 | 0.000394 ± 0.000171 |
| $\gamma$ Lep | UT 2009 Jan 10 | -0.96 | 55.7760 | 55.1121 | -0.000252 | 0.002443 | 0.002500 | 0.001990 ± 0.000564 |
| $\gamma$ Lep | UT 2009 Jan 13 | -1.25 | 56.3146 | 56.7235 | 0.000925 | 0.001408 | 0.002500 | 0.002068 ± 0.000586 |
| $\gamma$ Oph | UT 2008 Jul 16 | -1.09 | 56.0575 | 63.7328 | 0.016556 | 0.004257 | 0.003500 | 0.000306 ± 0.000173 |
| $\gamma$ Oph | UT 2008 Jul 16 | 0.93 | 44.6871 | 65.0191 | 0.007115 | 0.003479 | 0.003500 | 0.000264 ± 0.000149 |
| $\gamma$ Oph | UT 2008 Jul 16 | 2.18 | 31.1048 | 65.6124 | 0.009111 | 0.001849 | 0.003500 | 0.000224 ± 0.000127 |
| $\gamma$ Oph | UT 2008 Jul 17 | -1.20 | 56.2478 | 63.6559 | 0.001213 | 0.005165 | 0.003500 | 0.000306 ± 0.000173 |
| $\gamma$ Oph | UT 2008 Jul 17 | 1.12 | 42.8833 | 65.1234 | 0.001334 | 0.002470 | 0.003500 | 0.000258 ± 0.000146 |
| $\gamma$ Oph | UT 2008 Jul 17 | 2.20 | 30.9258 | 65.6180 | 0.010596 | 0.002324 | 0.003500 | 0.000223 ± 0.000126 |
| $\gamma$ Ser | UT 2008 Apr 16 | 0.73 | 46.4220 | 69.8053 | 0.001319 | 0.001941 | 0.003000 | 0.002031 ± 0.000880 |
| $\gamma$ Ser | UT 2008 Apr 16 | 1.46 | 39.4365 | 72.0252 | 0.008282 | 0.002438 | 0.003000 | 0.001948 ± 0.000844 |
| $\gamma$ Ser | UT 2008 Apr 16 | 2.62 | 25.5293 | 74.7012 | -0.000487 | 0.003023 | 0.003000 | 0.001801 ± 0.000780 |
| $\gamma$ Ser | UT 2008 Apr 17 | 2.40 | 28.4275 | 74.2770 | -0.006036 | 0.002157 | 0.003000 | 0.001828 ± 0.000792 |
| $\gamma$ Ser | UT 2008 Apr 17 | 2.93 | 21.3939 | 75.2099 | -0.008536 | 0.002857 | 0.003000 | 0.001767 ± 0.000766 |
| $\iota$ Peg | UT 2008 Jul 14 | 1.27 | 41.3608 | 73.8551 | 0.011249 | 0.001265 | 0.002500 | 0.015200 ± 0.000590 |
| $\iota$ Peg | UT 2008 Jul 14 | 1.98 | 33.5176 | 76.8496 | 0.011541 | 0.001314 | 0.002500 | 0.015300 ± 0.000579 |
| $\iota$ Per | UT 2009 Jan 11 | 1.34 | 40.6795 | 71.0095 | -0.003608 | 0.002332 | 0.002500 | 0.001968 ± 0.000390 |
| $\iota$ Psc | UT 2008 Aug 18 | -0.46 | 54.0668 | 64.7804 | -0.005178 | 0.003428 | 0.003000 | 0.001545 ± 0.000416 |
| $\iota$ Psc | UT 2008 Oct 13 | -1.79 | 56.4656 | 62.8581 | 0.003308 | 0.002344 | 0.003000 | 0.001550 ± 0.000417 |

Table 3—Continued

| Name | Date | HA | u (m) | v (m) | $L_{calibrated}$ | $\sigma^{formal}_{L_{calibrated}}$ | $\sigma^{ext.}_{L_{calibrated}}$ | $L_{\star}$ |
|---|---|---|---|---|---|---|---|---|
| $\iota$ Psc | UT 2008 Oct 13 | -0.93 | 55.7001 | 64.1049 | 0.003550 | 0.003045 | 0.003000 | 0.001565 ± 0.000421 |
| $\kappa$-1 Cet | UT 2009 Jan 10 | -0.13 | 52.4351 | 64.5998 | -0.000822 | 0.003682 | 0.003000 | 0.001681 ± 0.000489 |
| KX Lib | UT 2008 May 26 | 1.25 | 41.6389 | 45.0668 | 0.006615 | 0.002061 | 0.002500 | 0.001588 ± 0.001008 |
| KX Lib | UT 2008 May 26 | 2.14 | 31.5908 | 41.9053 | 0.002920 | 0.002855 | 0.002500 | 0.001162 ± 0.000737 |
| $\lambda$ Aur | UT 2009 Jan 12 | 0.62 | 47.3438 | 68.4446 | 0.007726 | 0.002093 | 0.003000 | 0.001390 ± 0.000862 |
| $NSV$ 4765 | UT 2009 Jan 10 | 1.90 | 34.5687 | 75.2646 | -0.004485 | 0.002523 | 0.003000 | 0.003350 ± 0.001718 |
| $\tau$ Boo | UT 2008 May 25 | 1.40 | 40.1023 | 72.4754 | 0.004442 | 0.003189 | 0.003000 | 0.000903 ± 0.000223 |
| $\tau$ Boo | UT 2008 May 25 | 2.28 | 29.8482 | 74.9182 | 0.005715 | 0.003108 | 0.003000 | 0.000856 ± 0.000211 |
| $\tau$ Boo | UT 2008 May 25 | 3.05 | 19.6580 | 76.4125 | 0.002326 | 0.002604 | 0.003000 | 0.000820 ± 0.000202 |
| $\tau$ Boo | UT 2008 May 27 | 1.84 | 35.2712 | 73.7720 | 0.004587 | 0.001592 | 0.003000 | 0.000880 ± 0.000217 |
| $\tau$ Boo | UT 2008 May 27 | 2.76 | 23.6261 | 75.9201 | 0.003603 | 0.002224 | 0.003000 | 0.000832 ± 0.000206 |
| $\theta$ Per | UT 2009 Jan 11 | 0.49 | 48.3812 | 63.5750 | 0.011956 | 0.001680 | 0.003000 | 0.001333 ± 0.000575 |
| $\theta$ Per | UT 2009 Jan 11 | 2.32 | 29.3767 | 78.0123 | -0.007308 | 0.002905 | 0.003000 | 0.001451 ± 0.000626 |
| $\theta$ Per | UT 2009 Jan 12 | 1.93 | 34.1275 | 75.5641 | -0.000798 | 0.001769 | 0.003000 | 0.001435 ± 0.000619 |
| $\upsilon$ And | UT 2008 Nov 12 | 2.53 | 26.6782 | 80.5610 | -0.004274 | 0.002613 | 0.002500 | 0.001845 ± 0.000425 |
| $\upsilon$ And | UT 2008 Nov 12 | 3.43 | 14.2844 | 83.7755 | 0.002662 | 0.001777 | 0.002500 | 0.001851 ± 0.000426 |

Note. — Table 3 is published in its entirety in the electronic edition of the Astrophysical Journal. A portion is shown here for guidance regarding its form and content.

Table 4. Number of zodis for each observation and each zodi disk orientation.

| Name | Date | HA | $i_{disk}$ (°) | PA$_{disk}$ (°) | $z_{ij}$ | $\sigma_{ij}^{formal}$ | $\sigma_{j}^{ext}$ | $\sigma^{\star}$ | $R_{\text{half-light}}$ (AU) |
|---|---|---|---|---|---|---|---|---|---|
| 107 Psc | UT 2008 Oct 13 | 2.54 | 0.0 | 109.0 | 256 | 88 | 147 | 27 | 0.07 |
| 107 Psc | UT 2008 Oct 13 | 2.54 | 90.0 | 109.0 | 358 | 123 | 205 | 38 | 0.07 |
| 107 Psc | UT 2008 Oct 13 | 2.54 | 90.0 | 19.0 | 244 | 84 | 141 | 26 | 0.06 |
| 107 Psc | UT 2008 Oct 13 | 3.27 | 0.0 | 102.0 | -98 | 100 | 148 | 26 | 0.07 |
| 107 Psc | UT 2008 Oct 13 | 3.27 | 90.0 | 102.0 | -135 | 137 | 202 | 36 | 0.07 |
| 107 Psc | UT 2008 Oct 13 | 3.27 | 90.0 | 12.0 | -95 | 97 | 142 | 25 | 0.06 |
| 1 Ori | UT 2008 Feb 17 | 0.41 | 0.0 | 125.2 | -143 | 37 | 67 | 21 | 0.10 |
| 1 Ori | UT 2008 Feb 17 | 0.41 | 90.0 | 125.2 | -180 | 47 | 84 | 27 | 0.10 |
| 1 Ori | UT 2008 Feb 17 | 0.41 | 90.0 | 35.2 | -113 | 29 | 53 | 17 | 0.08 |
| 1 Ori | UT 2008 Feb 17 | 1.10 | 0.0 | 120.8 | 1 | 25 | 67 | 21 | 0.10 |
| 1 Ori | UT 2008 Feb 17 | 1.10 | 90.0 | 120.8 | 1 | 31 | 83 | 26 | 0.10 |
| 1 Ori | UT 2008 Feb 17 | 1.10 | 90.0 | 30.8 | 0 | 20 | 53 | 17 | 0.08 |
| 1 Ori | UT 2008 Feb 17 | 1.83 | 0.0 | 115.3 | -90 | 42 | 67 | 20 | 0.10 |
| 1 Ori | UT 2008 Feb 17 | 1.83 | 90.0 | 115.3 | -105 | 50 | 78 | 24 | 0.10 |
| 1 Ori | UT 2008 Feb 17 | 1.83 | 90.0 | 25.3 | -76 | 36 | 57 | 17 | 0.08 |
| 1 Ori | UT 2008 Feb 18 | 0.10 | 0.0 | 127.0 | 158 | 60 | 67 | 22 | 0.10 |
| 1 Ori | UT 2008 Feb 18 | 0.10 | 90.0 | 127.0 | 189 | 71 | 79 | 26 | 0.10 |
| 1 Ori | UT 2008 Feb 18 | 0.10 | 90.0 | 37.0 | 132 | 50 | 56 | 18 | 0.08 |
| 1 Ori | UT 2008 Feb 18 | 1.22 | 0.0 | 120.0 | 121 | 89 | 67 | 21 | 0.10 |
| 1 Ori | UT 2008 Feb 18 | 1.22 | 90.0 | 120.0 | 150 | 109 | 82 | 26 | 0.10 |
| 1 Ori | UT 2008 Feb 18 | 1.22 | 90.0 | 30.0 | 98 | 72 | 54 | 17 | 0.08 |
| 47 UMa | UT 2009 Jan 10 | -1.08 | 0.0 | 136.5 | 62 | 154 | 172 | 9 | 0.14 |
| 47 UMa | UT 2009 Jan 10 | -1.08 | 90.0 | 136.5 | 84 | 208 | 233 | 13 | 0.14 |
| 47 UMa | UT 2009 Jan 10 | -1.08 | 90.0 | 46.5 | 61 | 151 | 169 | 9 | 0.11 |
| 61 Cyg A | UT 2008 Aug 17 | 0.46 | 0.0 | 125.7 | 367 | 173 | 165 | 184 | -0.01 |
| 61 Cyg A | UT 2008 Aug 17 | 0.46 | 90.0 | 125.7 | 507 | 237 | 226 | 254 | 0.03 |
| 61 Cyg A | UT 2008 Aug 17 | 0.46 | 90.0 | 35.7 | 324 | 153 | 146 | 162 | 0.03 |
| 61 Cyg A | UT 2008 Aug 17 | 1.12 | 0.0 | 120.6 | -54 | 151 | 165 | 187 | 0.03 |
| 61 Cyg A | UT 2008 Aug 17 | 1.12 | 90.0 | 120.6 | -75 | 210 | 229 | 261 | 0.03 |
| 61 Cyg A | UT 2008 Aug 17 | 1.12 | 90.0 | 30.6 | -47 | 131 | 143 | 163 | 0.03 |
| 61 Cyg A | UT 2008 Aug 17 | 1.70 | 0.0 | 115.7 | 104 | 206 | 165 | 190 | 0.03 |

Table 4—Continued

| Name | Date | HA | $i_{disk}$ (°) | PA$_{disk}$ (°) | $z_{ij}$ | $\sigma_{ij}^{formal}$ | $\sigma_{j}^{ext}$ | $\sigma^{\star}$ | $R_{\text{half-light}}$ (AU) |
|---|---|---|---|---|---|---|---|---|---|
| 61 Cyg A | UT 2008 Aug 17 | 1.70 | 90.0 | 115.7 | 143 | 282 | 226 | 262 | 0.03 |
| 61 Cyg A | UT 2008 Aug 17 | 1.70 | 90.0 | 25.7 | 91 | 180 | 144 | 166 | 0.03 |
| 70 Oph | UT 2008 Aug 17 | 1.27 | 0.0 | 122.5 | 551 | 129 | 160 | 134 | 0.06 |
| 70 Oph | UT 2008 Aug 17 | 1.27 | 90.0 | 122.5 | 755 | 176 | 218 | 184 | 0.06 |
| 70 Oph | UT 2008 Aug 17 | 1.27 | 90.0 | 32.5 | 461 | 108 | 134 | 113 | 0.05 |
| 70 Oph | UT 2008 Aug 17 | 1.86 | 0.0 | 118.1 | 42 | 147 | 161 | 124 | 0.06 |
| 70 Oph | UT 2008 Aug 17 | 1.86 | 90.0 | 118.1 | 56 | 195 | 214 | 165 | 0.06 |
| 70 Oph | UT 2008 Aug 17 | 1.86 | 90.0 | 28.1 | 36 | 126 | 139 | 107 | 0.05 |
| 70 Oph | UT 2008 Aug 18 | 0.35 | 0.0 | 127.4 | -223 | 174 | 159 | 147 | 0.06 |
| 70 Oph | UT 2008 Aug 18 | 0.35 | 90.0 | 127.4 | -294 | 229 | 210 | 194 | 0.06 |
| 70 Oph | UT 2008 Aug 18 | 0.35 | 90.0 | 37.4 | -190 | 148 | 136 | 126 | -0.01 |
| 70 Oph | UT 2008 Aug 18 | 1.03 | 0.0 | 123.9 | -430 | 893 | 160 | 137 | 0.06 |
| 70 Oph | UT 2008 Aug 18 | 1.03 | 90.0 | 123.9 | -585 | 1216 | 218 | 186 | 0.06 |
| 70 Oph | UT 2008 Aug 18 | 1.03 | 90.0 | 33.9 | -358 | 743 | 133 | 114 | 0.05 |
| 70 Oph | UT 2008 Aug 18 | 1.89 | 0.0 | 118.0 | 140 | 214 | 161 | 123 | 0.06 |
| 70 Oph | UT 2008 Aug 18 | 1.89 | 90.0 | 118.0 | 187 | 284 | 214 | 165 | 0.06 |
| 70 Oph | UT 2008 Aug 18 | 1.89 | 90.0 | 28.0 | 121 | 184 | 139 | 106 | 0.05 |
| HIP 54035 | UT 2008 Apr 14 | 2.75 | 0.0 | 106.2 | -286 | 191 | 207 | 110 | 0.02 |
| HIP 54035 | UT 2008 Apr 14 | 2.75 | 90.0 | 106.2 | -456 | 305 | 331 | 176 | 0.02 |
| HIP 54035 | UT 2008 Apr 14 | 2.75 | 90.0 | 16.2 | -276 | 184 | 200 | 106 | -0.01 |
| HIP 54035 | UT 2008 Apr 14 | 3.45 | 0.0 | 99.5 | -302 | 197 | 207 | 110 | 0.02 |
| HIP 54035 | UT 2008 Apr 14 | 3.45 | 90.0 | 99.5 | -473 | 309 | 325 | 172 | -0.01 |
| HIP 54035 | UT 2008 Apr 14 | 3.45 | 90.0 | 9.5 | -293 | 191 | 201 | 107 | -0.01 |
| HIP 54035 | UT 2009 Jan 10 | -0.23 | 0.0 | 130.1 | -101 | 172 | 210 | 105 | -0.01 |
| HIP 54035 | UT 2009 Jan 10 | -0.23 | 90.0 | 130.1 | -164 | 278 | 338 | 171 | 0.02 |
| HIP 54035 | UT 2009 Jan 10 | -0.23 | 90.0 | 40.1 | -98 | 168 | 204 | 103 | -0.01 |
| $\alpha$ Aql | UT 2008 May 25 | -1.22 | 0.0 | 131.5 | 828 | 64 | 182 | 15 | 0.17 |
| $\alpha$ Aql | UT 2008 May 25 | -1.22 | 90.0 | 131.5 | 462 | 35 | 101 | 8 | 0.11 |
| $\alpha$ Aql | UT 2008 May 25 | -1.22 | 90.0 | 41.5 | 766 | 59 | 168 | 14 | 0.08 |
| $\alpha$ Aql | UT 2008 May 25 | -0.42 | 0.0 | 129.5 | 517 | 74 | 182 | 15 | 0.17 |
| $\alpha$ Aql | UT 2008 May 25 | -0.42 | 90.0 | 129.5 | 312 | 45 | 109 | 9 | 0.11 |

Table 4—Continued

| Name | Date | HA | $i_{disk}$ (°) | $\mathrm{PA}_{disk}$ (°) | $z_{ij}$ | $\sigma_{ij}^{formal}$ | $\sigma_j^{ext}$ | $\sigma^\star$ | $R_{\text{half-light}}$ (AU) |
|---|---|---|---|---|---|---|---|---|---|
| $\alpha$ Aql | UT 2008 May 25 | -0.42 | 90.0 | 39.5 | 449 | 64 | 158 | 13 | 0.08 |
| $\alpha$ Aql | UT 2008 May 26 | -0.93 | 0.0 | 130.9 | 579 | 99 | 182 | 15 | 0.17 |
| $\alpha$ Aql | UT 2008 May 26 | -0.93 | 90.0 | 130.9 | 319 | 54 | 100 | 8 | 0.11 |
| $\alpha$ Aql | UT 2008 May 26 | -0.93 | 90.0 | 40.9 | 539 | 92 | 169 | 14 | 0.08 |
| $\alpha$ Aql | UT 2008 May 26 | -0.25 | 0.0 | 128.9 | 1211 | 105 | 182 | 15 | 0.17 |
| $\alpha$ Aql | UT 2008 May 26 | -0.25 | 90.0 | 128.9 | 809 | 70 | 121 | 10 | 0.10 |
| $\alpha$ Aql | UT 2008 May 26 | -0.25 | 90.0 | 38.9 | 971 | 84 | 146 | 12 | 0.09 |
| $\alpha$ Aql | UT 2008 May 26 | 0.35 | 0.0 | 126.4 | 132 | 72 | 182 | 14 | 0.17 |
| $\alpha$ Aql | UT 2008 May 26 | 0.35 | 90.0 | 126.4 | 128 | 69 | 175 | 14 | 0.10 |
| $\alpha$ Aql | UT 2008 May 26 | 0.35 | 90.0 | 36.4 | 72 | 39 | 99 | 8 | 0.09 |
| $\beta$ Com | UT 2008 Feb 16 | 0.20 | 0.0 | 126.5 | 568 | 114 | 131 | 14 | -0.02 |
| $\beta$ Com | UT 2008 Feb 16 | 0.20 | 90.0 | 126.5 | 783 | 155 | 179 | 19 | 0.09 |
| $\beta$ Com | UT 2008 Feb 16 | 0.20 | 90.0 | 36.5 | 496 | 99 | 114 | 12 | 0.07 |
| $\beta$ Com | UT 2008 Feb 16 | 1.08 | 0.0 | 120.6 | 35 | 140 | 131 | 14 | 0.09 |
| $\beta$ Com | UT 2008 Feb 16 | 1.08 | 90.0 | 120.6 | 48 | 189 | 177 | 19 | 0.09 |
| $\beta$ Com | UT 2008 Feb 16 | 1.08 | 90.0 | 30.6 | 31 | 123 | 115 | 12 | 0.07 |
| $\beta$ Com | UT 2008 Feb 16 | 1.87 | 0.0 | 114.4 | 92 | 88 | 131 | 13 | 0.09 |
| $\beta$ Com | UT 2008 Feb 16 | 1.87 | 90.0 | 114.4 | 120 | 116 | 171 | 18 | 0.09 |
| $\beta$ Com | UT 2008 Feb 16 | 1.87 | 90.0 | 24.4 | 83 | 80 | 118 | 12 | 0.07 |
| $\beta$ Vir | UT 2008 Feb 17 | 1.66 | 0.0 | 120.0 | -201 | 139 | 111 | 26 | 0.13 |
| $\beta$ Vir | UT 2008 Feb 17 | 1.66 | 90.0 | 120.0 | -264 | 183 | 147 | 34 | 0.13 |
| $\beta$ Vir | UT 2008 Feb 17 | 1.66 | 90.0 | 30.0 | -168 | 116 | 93 | 22 | 0.11 |
| $\beta$ Vir | UT 2008 Feb 17 | 2.33 | 0.0 | 114.3 | 261 | 157 | 112 | 24 | 0.15 |
| $\beta$ Vir | UT 2008 Feb 17 | 2.33 | 90.0 | 114.3 | 335 | 200 | 143 | 31 | 0.14 |
| $\beta$ Vir | UT 2008 Feb 17 | 2.33 | 90.0 | 24.3 | 227 | 136 | 97 | 21 | 0.11 |
| $\beta$ Vir | UT 2008 Feb 17 | 3.04 | 0.0 | 107.0 | 431 | 226 | 112 | 22 | 0.15 |
| $\beta$ Vir | UT 2008 Feb 17 | 3.04 | 90.0 | 107.0 | 532 | 276 | 138 | 27 | 0.14 |
| $\beta$ Vir | UT 2008 Feb 17 | 3.04 | 90.0 | 17.0 | 390 | 204 | 102 | 20 | 0.11 |
| $\beta$ Vir | UT 2008 Feb 18 | 1.55 | 0.0 | 120.7 | -343 | 284 | 111 | 26 | 0.13 |
| $\beta$ Vir | UT 2008 Feb 18 | 1.55 | 90.0 | 120.7 | -453 | 376 | 147 | 35 | 0.13 |
| $\beta$ Vir | UT 2008 Feb 18 | 1.55 | 90.0 | 30.7 | -286 | 236 | 93 | 22 | 0.11 |

Table 4—Continued

| Name | Date | HA | $i_{disk}$ (°) | PA$_{disk}$ (°) | $z_{ij}$ | $\sigma_{ij}^{formal}$ | $\sigma_j^{ext}$ | $\sigma^{\star}$ | $R_{\text{half-light}}$ (AU) |
|---|---|---|---|---|---|---|---|---|---|
| $\beta$ Vir | UT 2008 Feb 18 | 2.56 | 0.0 | 112.1 | -542 | 231 | 112 | 23 | 0.15 |
| $\beta$ Vir | UT 2008 Feb 18 | 2.56 | 90.0 | 112.1 | -678 | 291 | 141 | 29 | 0.14 |
| $\beta$ Vir | UT 2008 Feb 18 | 2.56 | 90.0 | 22.1 | -477 | 203 | 99 | 20 | 0.11 |
| $\beta$ Vir | UT 2008 Feb 18 | 2.69 | 0.0 | 110.7 | 20 | 367 | 112 | 23 | 0.15 |
| $\beta$ Vir | UT 2008 Feb 18 | 2.69 | 90.0 | 110.7 | 26 | 459 | 140 | 28 | 0.14 |
| $\beta$ Vir | UT 2008 Feb 18 | 2.69 | 90.0 | 20.7 | 18 | 326 | 100 | 20 | 0.11 |
| $\beta$ Vir | UT 2008 Feb 18 | 3.25 | 0.0 | 104.6 | 221 | 209 | 113 | 21 | 0.15 |
| $\beta$ Vir | UT 2008 Feb 18 | 3.25 | 90.0 | 104.6 | 268 | 253 | 136 | 26 | 0.15 |
| $\beta$ Vir | UT 2008 Feb 18 | 3.25 | 90.0 | 14.6 | 201 | 191 | 103 | 19 | 0.11 |
| $\chi$-1 Ori | UT 2009 Jan 10 | 1.15 | 0.0 | 120.4 | 368 | 88 | 146 | 28 | 0.09 |
| $\chi$-1 Ori | UT 2009 Jan 10 | 1.15 | 90.0 | 120.4 | 483 | 115 | 190 | 37 | 0.09 |
| $\chi$-1 Ori | UT 2009 Jan 10 | 1.15 | 90.0 | 30.4 | 337 | 81 | 133 | 25 | 0.07 |
| $\chi$-1 Ori | UT 2009 Jan 10 | 1.90 | 0.0 | 114.7 | -67 | 101 | 146 | 27 | 0.09 |
| $\chi$-1 Ori | UT 2009 Jan 10 | 1.90 | 90.0 | 114.7 | -87 | 130 | 188 | 34 | 0.09 |
| $\chi$-1 Ori | UT 2009 Jan 10 | 1.90 | 90.0 | 24.7 | -62 | 94 | 135 | 25 | 0.07 |
| $\chi$-1 Ori | UT 2009 Jan 13 | 2.35 | 0.0 | 110.9 | 4 | 121 | 146 | 26 | -0.02 |
| $\chi$-1 Ori | UT 2009 Jan 13 | 2.35 | 90.0 | 110.9 | 5 | 154 | 186 | 33 | 0.09 |
| $\chi$-1 Ori | UT 2009 Jan 13 | 2.35 | 90.0 | 20.9 | 4 | 113 | 136 | 24 | 0.07 |
| $\chi$-1 Ori | UT 2009 Jan 13 | 3.35 | 0.0 | 101.2 | -281 | 100 | 146 | 25 | -0.02 |
| $\chi$-1 Ori | UT 2009 Jan 13 | 3.35 | 90.0 | 101.2 | -348 | 124 | 182 | 31 | 0.09 |
| $\chi$-1 Ori | UT 2009 Jan 13 | 3.35 | 90.0 | 11.2 | -267 | 95 | 139 | 24 | -0.02 |
| $\delta$ Tri | UT 2008 Nov 13 | 0.44 | 0.0 | 125.4 | -589 | 283 | 161 | 26 | 0.11 |
| $\delta$ Tri | UT 2008 Nov 13 | 0.44 | 90.0 | 125.4 | -840 | 408 | 232 | 37 | 0.11 |
| $\delta$ Tri | UT 2008 Nov 13 | 0.44 | 90.0 | 35.4 | -531 | 256 | 146 | 23 | 0.09 |
| $\delta$ Tri | UT 2008 Nov 13 | 1.95 | 0.0 | 113.5 | -272 | 163 | 161 | 26 | 0.11 |
| $\delta$ Tri | UT 2008 Nov 13 | 1.95 | 90.0 | 113.5 | -381 | 229 | 226 | 37 | 0.11 |
| $\delta$ Tri | UT 2008 Nov 13 | 1.95 | 90.0 | 23.5 | -249 | 149 | 147 | 24 | 0.09 |
| $\eta$ Crv | UT 2008 Apr 17 | -1.93 | 0.0 | 132.1 | 1145 | 193 | 140 | 20 | -0.04 |
| $\eta$ Crv | UT 2008 Apr 17 | -1.93 | 90.0 | 132.1 | 1378 | 228 | 165 | 24 | 0.20 |
| $\eta$ Crv | UT 2008 Apr 17 | -1.93 | 90.0 | 42.1 | 1121 | 189 | 137 | 19 | 0.15 |
| $\eta$ Crv | UT 2008 May 24 | 0.71 | 0.0 | 131.9 | 1446 | 134 | 144 | 14 | 0.22 |

Table 4—Continued

| Name | Date | HA | $i_{disk}$ (°) | PA$_{disk}$ (°) | $z_{ij}$ | $\sigma_{ij}^{formal}$ | $\sigma_j^{ext}$ | $\sigma^\star$ | $R_{\text{half-light}}$ (AU) |
|---|---|---|---|---|---|---|---|---|---|
| $\eta$ Crv | UT 2008 May 24 | 0.71 | 90.0 | 131.9 | 2131 | 192 | 205 | 21 | 0.22 |
| $\eta$ Crv | UT 2008 May 24 | 0.71 | 90.0 | 41.9 | 1284 | 119 | 127 | 12 | 0.18 |
| $\eta$ Crv | UT 2008 May 24 | 1.91 | 0.0 | 125.4 | 950 | 166 | 148 | 10 | 0.25 |
| $\eta$ Crv | UT 2008 May 24 | 1.91 | 90.0 | 125.4 | 1516 | 259 | 230 | 17 | 0.25 |
| $\eta$ Crv | UT 2008 May 24 | 1.91 | 90.0 | 35.4 | 818 | 143 | 127 | 9 | 0.18 |
| $\eta$ Crv | UT 2008 May 24 | 2.61 | 0.0 | 118.7 | 648 | 484 | 151 | 8 | 0.27 |
| $\eta$ Crv | UT 2008 May 24 | 2.61 | 90.0 | 118.7 | 1050 | 770 | 240 | 14 | 0.25 |
| $\eta$ Crv | UT 2008 May 24 | 2.61 | 90.0 | 28.7 | 564 | 421 | 131 | 7 | 0.22 |
| $\gamma$ Lep | UT 2009 Jan 10 | -0.96 | 0.0 | 135.3 | -105 | 114 | 117 | 26 | 0.11 |
| $\gamma$ Lep | UT 2009 Jan 10 | -0.96 | 90.0 | 135.3 | -117 | 128 | 131 | 29 | 0.12 |
| $\gamma$ Lep | UT 2009 Jan 10 | -0.96 | 90.0 | 45.3 | -99 | 108 | 111 | 25 | 0.09 |
| $\gamma$ Lep | UT 2009 Jan 13 | -1.25 | 0.0 | 134.8 | -53 | 66 | 117 | 27 | 0.11 |
| $\gamma$ Lep | UT 2009 Jan 13 | -1.25 | 90.0 | 134.8 | -58 | 72 | 128 | 30 | 0.12 |
| $\gamma$ Lep | UT 2009 Jan 13 | -1.25 | 90.0 | 44.8 | -51 | 63 | 112 | 26 | -0.02 |
| $\gamma$ Oph | UT 2008 Jul 16 | -1.09 | 0.0 | 131.3 | 382 | 98 | 81 | 4 | 0.35 |
| $\gamma$ Oph | UT 2008 Jul 16 | -1.09 | 90.0 | 131.3 | 410 | 104 | 86 | 4 | 0.35 |
| $\gamma$ Oph | UT 2008 Jul 16 | -1.09 | 90.0 | 41.3 | 374 | 96 | 79 | 3 | 0.23 |
| $\gamma$ Oph | UT 2008 Jul 16 | 0.93 | 0.0 | 124.5 | 160 | 81 | 81 | 3 | 0.35 |
| $\gamma$ Oph | UT 2008 Jul 16 | 0.93 | 90.0 | 124.5 | 218 | 109 | 110 | 4 | 0.35 |
| $\gamma$ Oph | UT 2008 Jul 16 | 0.93 | 90.0 | 34.5 | 132 | 66 | 67 | 2 | 0.29 |
| $\gamma$ Oph | UT 2008 Jul 16 | 2.18 | 0.0 | 115.4 | 211 | 43 | 82 | 3 | 0.35 |
| $\gamma$ Oph | UT 2008 Jul 16 | 2.18 | 90.0 | 115.4 | 274 | 56 | 106 | 3 | 0.38 |
| $\gamma$ Oph | UT 2008 Jul 16 | 2.18 | 90.0 | 25.4 | 184 | 38 | 71 | 2 | 0.29 |
| $\gamma$ Oph | UT 2008 Jul 17 | -1.20 | 0.0 | 131.5 | 21 | 119 | 81 | 4 | 0.35 |
| $\gamma$ Oph | UT 2008 Jul 17 | -1.20 | 90.0 | 131.5 | 22 | 126 | 85 | 4 | 0.35 |
| $\gamma$ Oph | UT 2008 Jul 17 | -1.20 | 90.0 | 41.5 | 20 | 117 | 79 | 3 | 0.23 |
| $\gamma$ Oph | UT 2008 Jul 17 | 1.12 | 0.0 | 123.4 | 25 | 57 | 81 | 3 | 0.35 |
| $\gamma$ Oph | UT 2008 Jul 17 | 1.12 | 90.0 | 123.4 | 33 | 77 | 110 | 4 | 0.35 |
| $\gamma$ Oph | UT 2008 Jul 17 | 1.12 | 90.0 | 33.4 | 20 | 47 | 67 | 2 | 0.29 |
| $\gamma$ Oph | UT 2008 Jul 17 | 2.20 | 0.0 | 115.2 | 246 | 54 | 82 | 3 | 0.35 |
| $\gamma$ Oph | UT 2008 Jul 17 | 2.20 | 90.0 | 115.2 | 320 | 70 | 106 | 3 | 0.38 |

Table 4—Continued

| Name | Date | HA | $i_{disk}$ (°) | PA$_{disk}$ (°) | $z_{ij}$ | $\sigma_{ij}^{formal}$ | $\sigma_j^{ext}$ | $\sigma^\star$ | $R_{\text{half-light}}$ (AU) |
|---|---|---|---|---|---|---|---|---|---|
| $\gamma$ Oph | UT 2008 Jul 17 | 2.20 | 90.0 | 25.2 | 215 | 47 | 71 | 2 | 0.29 |
| $\gamma$ Ser | UT 2008 Apr 16 | 0.73 | 0.0 | 123.6 | -25 | 69 | 106 | 31 | 0.13 |
| $\gamma$ Ser | UT 2008 Apr 16 | 0.73 | 90.0 | 123.6 | -33 | 89 | 138 | 40 | 0.13 |
| $\gamma$ Ser | UT 2008 Apr 16 | 0.73 | 90.0 | 33.6 | -21 | 59 | 92 | 27 | 0.09 |
| $\gamma$ Ser | UT 2008 Apr 16 | 1.46 | 0.0 | 118.7 | 227 | 86 | 107 | 30 | 0.13 |
| $\gamma$ Ser | UT 2008 Apr 16 | 1.46 | 90.0 | 118.7 | 287 | 108 | 134 | 38 | 0.13 |
| $\gamma$ Ser | UT 2008 Apr 16 | 1.46 | 90.0 | 28.7 | 203 | 77 | 95 | 27 | 0.09 |
| $\gamma$ Ser | UT 2008 Apr 16 | 2.62 | 0.0 | 108.9 | -81 | 108 | 107 | 27 | 0.13 |
| $\gamma$ Ser | UT 2008 Apr 16 | 2.62 | 90.0 | 108.9 | -98 | 130 | 129 | 33 | 0.13 |
| $\gamma$ Ser | UT 2008 Apr 16 | 2.62 | 90.0 | 18.9 | -75 | 100 | 99 | 25 | -0.02 |
| $\gamma$ Ser | UT 2008 Apr 17 | 2.40 | 0.0 | 110.9 | -279 | 77 | 107 | 28 | 0.13 |
| $\gamma$ Ser | UT 2008 Apr 17 | 2.40 | 90.0 | 110.9 | -338 | 93 | 130 | 34 | 0.13 |
| $\gamma$ Ser | UT 2008 Apr 17 | 2.40 | 90.0 | 20.9 | -257 | 70 | 98 | 25 | -0.02 |
| $\gamma$ Ser | UT 2008 Apr 17 | 2.93 | 0.0 | 105.9 | -366 | 102 | 107 | 27 | 0.13 |
| $\gamma$ Ser | UT 2008 Apr 17 | 2.93 | 90.0 | 105.9 | -435 | 122 | 128 | 32 | 0.13 |
| $\gamma$ Ser | UT 2008 Apr 17 | 2.93 | 90.0 | 15.9 | -340 | 95 | 100 | 25 | -0.02 |
| $\iota$ Peg | UT 2008 Jul 14 | 1.27 | 0.0 | 119.2 | -167 | 53 | 105 | 25 | 0.14 |
| $\iota$ Peg | UT 2008 Jul 14 | 1.27 | 90.0 | 119.2 | -212 | 66 | 131 | 31 | 0.14 |
| $\iota$ Peg | UT 2008 Jul 14 | 1.27 | 90.0 | 29.2 | -147 | 46 | 92 | 21 | -0.02 |
| $\iota$ Peg | UT 2008 Jul 14 | 1.98 | 0.0 | 113.6 | -159 | 55 | 105 | 24 | 0.14 |
| $\iota$ Peg | UT 2008 Jul 14 | 1.98 | 90.0 | 113.6 | -196 | 67 | 128 | 30 | 0.14 |
| $\iota$ Peg | UT 2008 Jul 14 | 1.98 | 90.0 | 23.6 | -143 | 49 | 94 | 22 | -0.02 |
| $\iota$ Per | UT 2009 Jan 11 | 1.34 | 0.0 | 119.8 | -267 | 112 | 120 | 18 | 0.13 |
| $\iota$ Per | UT 2009 Jan 11 | 1.34 | 90.0 | 119.8 | -364 | 153 | 164 | 25 | 0.13 |
| $\iota$ Per | UT 2009 Jan 11 | 1.34 | 90.0 | 29.8 | -226 | 95 | 101 | 15 | 0.08 |
| $\iota$ Psc | UT 2008 Aug 18 | -0.46 | 0.0 | 129.8 | -303 | 155 | 136 | 18 | 0.17 |
| $\iota$ Psc | UT 2008 Aug 18 | -0.46 | 90.0 | 129.8 | -359 | 185 | 162 | 22 | 0.17 |
| $\iota$ Psc | UT 2008 Aug 18 | -0.46 | 90.0 | 39.8 | -289 | 148 | 129 | 17 | 0.11 |
| $\iota$ Psc | UT 2008 Oct 13 | -1.79 | 0.0 | 131.9 | 80 | 106 | 136 | 18 | 0.17 |
| $\iota$ Psc | UT 2008 Oct 13 | -1.79 | 90.0 | 131.9 | 92 | 122 | 156 | 21 | 0.17 |
| $\iota$ Psc | UT 2008 Oct 13 | -1.79 | 90.0 | 41.9 | 77 | 103 | 132 | 18 | 0.11 |

Table 4—Continued

| Name | Date | HA | $i_{disk}$ (°) | PA$_{disk}$ (°) | $z_{ij}$ | $\sigma_{ij}^{formal}$ | $\sigma_{j}^{ext}$ | $\sigma^{\star}$ | $R_{\text{half-light}}$ (AU) |
|---|---|---|---|---|---|---|---|---|---|
| $\iota$ Psc | UT 2008 Oct 13 | -0.93 | 0.0 | 131.0 | 90 | 138 | 135 | 19 | 0.17 |
| $\iota$ Psc | UT 2008 Oct 13 | -0.93 | 90.0 | 131.0 | 103 | 158 | 155 | 22 | 0.17 |
| $\iota$ Psc | UT 2008 Oct 13 | -0.93 | 90.0 | 41.0 | 88 | 134 | 132 | 18 | 0.11 |
| $\kappa$-1 Cet | UT 2009 Jan 10 | -0.13 | 0.0 | 129.1 | -107 | 158 | 129 | 21 | 0.09 |
| $\kappa$-1 Cet | UT 2009 Jan 10 | -0.13 | 90.0 | 129.1 | -145 | 213 | 174 | 28 | 0.09 |
| $\kappa$-1 Cet | UT 2009 Jan 10 | -0.13 | 90.0 | 39.1 | -101 | 148 | 121 | 19 | 0.07 |
| KX Lib | UT 2008 May 26 | 1.25 | 0.0 | 132.7 | 527 | 214 | 260 | 105 | 0.07 |
| KX Lib | UT 2008 May 26 | 1.25 | 90.0 | 132.7 | 871 | 350 | 425 | 174 | 0.07 |
| KX Lib | UT 2008 May 26 | 1.25 | 90.0 | 42.7 | 480 | 195 | 237 | 96 | 0.06 |
| KX Lib | UT 2008 May 26 | 2.14 | 0.0 | 127.0 | 191 | 309 | 270 | 80 | 0.08 |
| KX Lib | UT 2008 May 26 | 2.14 | 90.0 | 127.0 | 338 | 544 | 476 | 141 | 0.08 |
| KX Lib | UT 2008 May 26 | 2.14 | 90.0 | 37.0 | 171 | 278 | 243 | 72 | 0.06 |
| $\lambda$ Aur | UT 2009 Jan 12 | 0.62 | 0.0 | 124.7 | 341 | 111 | 160 | 46 | 0.13 |
| $\lambda$ Aur | UT 2009 Jan 12 | 0.62 | 90.0 | 124.7 | 487 | 158 | 227 | 66 | 0.13 |
| $\lambda$ Aur | UT 2009 Jan 12 | 0.62 | 90.0 | 34.7 | 304 | 99 | 143 | 41 | 0.10 |
| *NSV* 4765 | UT 2009 Jan 10 | 1.90 | 0.0 | 114.7 | -506 | 163 | 195 | 111 | 0.05 |
| *NSV* 4765 | UT 2009 Jan 10 | 1.90 | 90.0 | 114.7 | -775 | 252 | 300 | 170 | 0.04 |
| *NSV* 4765 | UT 2009 Jan 10 | 1.90 | 90.0 | 24.7 | -467 | 151 | 180 | 102 | 0.04 |
| $\tau$ Boo | UT 2008 May 25 | 1.40 | 0.0 | 119.0 | 155 | 139 | 131 | 9 | 0.16 |
| $\tau$ Boo | UT 2008 May 25 | 1.40 | 90.0 | 119.0 | 207 | 185 | 174 | 13 | 0.16 |
| $\tau$ Boo | UT 2008 May 25 | 1.40 | 90.0 | 29.0 | 143 | 129 | 121 | 9 | 0.12 |
| $\tau$ Boo | UT 2008 May 25 | 2.28 | 0.0 | 111.7 | 214 | 136 | 131 | 9 | 0.16 |
| $\tau$ Boo | UT 2008 May 25 | 2.28 | 90.0 | 111.7 | 281 | 178 | 171 | 12 | 0.16 |
| $\tau$ Boo | UT 2008 May 25 | 2.28 | 90.0 | 21.7 | 201 | 128 | 123 | 8 | 0.12 |
| $\tau$ Boo | UT 2008 May 25 | 3.05 | 0.0 | 104.4 | 66 | 114 | 132 | 8 | 0.16 |
| $\tau$ Boo | UT 2008 May 25 | 3.05 | 90.0 | 104.4 | 85 | 147 | 169 | 11 | 0.16 |
| $\tau$ Boo | UT 2008 May 25 | 3.05 | 90.0 | 14.4 | 63 | 109 | 125 | 8 | 0.12 |
| $\tau$ Boo | UT 2008 May 27 | 1.84 | 0.0 | 115.6 | 163 | 69 | 131 | 9 | 0.16 |
| $\tau$ Boo | UT 2008 May 27 | 1.84 | 90.0 | 115.6 | 215 | 91 | 172 | 12 | 0.16 |
| $\tau$ Boo | UT 2008 May 27 | 1.84 | 90.0 | 25.6 | 152 | 65 | 122 | 8 | 0.12 |
| $\tau$ Boo | UT 2008 May 27 | 2.76 | 0.0 | 107.3 | 122 | 97 | 132 | 9 | 0.16 |

Table 4—Continued

| Name | Date | HA | $i_{disk}$ (°) | PA$_{disk}$ (°) | $z_{ij}$ | $\sigma_{ij}^{formal}$ | $\sigma_j^{ext}$ | $\sigma^{\star}$ | $R_{\text{half-light}}$ (AU) |
|---|---|---|---|---|---|---|---|---|---|
| $\tau$ Boo | UT 2008 May 27 | 2.76 | 90.0 | 107.3 | 158 | 126 | 170 | 11 | 0.16 |
| $\tau$ Boo | UT 2008 May 27 | 2.76 | 90.0 | 17.3 | 115 | 92 | 124 | 8 | 0.12 |
| $\theta$ Per | UT 2009 Jan 11 | 0.49 | 0.0 | 127.3 | 400 | 62 | 111 | 21 | 0.13 |
| $\theta$ Per | UT 2009 Jan 11 | 0.49 | 90.0 | 127.3 | 544 | 84 | 150 | 29 | 0.13 |
| $\theta$ Per | UT 2009 Jan 11 | 0.49 | 90.0 | 37.3 | 352 | 55 | 98 | 19 | 0.09 |
| $\theta$ Per | UT 2009 Jan 11 | 2.32 | 0.0 | 110.6 | -321 | 107 | 111 | 22 | -0.02 |
| $\theta$ Per | UT 2009 Jan 11 | 2.32 | 90.0 | 110.6 | -436 | 146 | 151 | 31 | 0.11 |
| $\theta$ Per | UT 2009 Jan 11 | 2.32 | 90.0 | 20.6 | -276 | 92 | 95 | 19 | 0.09 |
| $\theta$ Per | UT 2009 Jan 12 | 1.93 | 0.0 | 114.3 | -82 | 65 | 111 | 22 | 0.13 |
| $\theta$ Per | UT 2009 Jan 12 | 1.93 | 90.0 | 114.3 | -113 | 90 | 152 | 31 | 0.11 |
| $\theta$ Per | UT 2009 Jan 12 | 1.93 | 90.0 | 24.3 | -70 | 56 | 95 | 19 | 0.09 |
| $\upsilon$ And | UT 2008 Nov 12 | 2.53 | 0.0 | 108.3 | -300 | 128 | 123 | 20 | 0.13 |
| $\upsilon$ And | UT 2008 Nov 12 | 2.53 | 90.0 | 108.3 | -394 | 169 | 162 | 27 | 0.13 |
| $\upsilon$ And | UT 2008 Nov 12 | 2.53 | 90.0 | 18.3 | -263 | 113 | 108 | 18 | 0.11 |
| $\upsilon$ And | UT 2008 Nov 12 | 3.43 | 0.0 | 99.7 | 40 | 87 | 123 | 21 | 0.13 |
| $\upsilon$ And | UT 2008 Nov 12 | 3.43 | 90.0 | 99.7 | 50 | 110 | 156 | 26 | 0.13 |
| $\upsilon$ And | UT 2008 Nov 12 | 3.43 | 90.0 | 9.7 | 36 | 79 | 111 | 18 | 0.11 |

Note. — Table 4 is published in its entirety in the electronic edition of the Astrophysical Journal. A portion is shown here for guidance regarding its form and content.

Table 5. Average number of zodis for each target.

| Name | $z \pm \sigma_z$ | $\chi = \frac{z}{\sigma_z}$ | $3\sigma$ Upper limits |
|---|---|---|---|
| | | Detections: | |
| $\eta$ Crv | 1246 ± 257 | 4.8 | ··· |
| | | Possible detections: | |
| $\gamma$ Oph | 198 ± 77 | 2.6 | 429 |
| $\alpha$ Aql | 573 ± 191 | 3.0 | 1146 |
| | | Non–detections: | |
| 107 Psc | 107 ± 192 | 0.6 | 683 |
| 1 Ori | 43 ± 48 | 0.9 | 187 |
| 47 UMa | 67 ± 187 | 0.4 | 628 |
| 61 Cyg A | 143 ± 194 | 0.7 | 725 |
| 70 Oph | 67 ± 159 | 0.4 | 544 |
| HIP 54035 | -227 ± 179 | -1.3 | 537 |
| $\beta$ Com | 237 ± 245 | 1.0 | 972 |
| $\beta$ Vir | -9 ± 214 | -0.0 | 642 |
| $\chi$-1 Ori | -60 ± 128 | -0.5 | 384 |
| $\delta$ Tri | -380 ± 191 | -2.0 | 573 |
| $\gamma$ Lep | -80 ± 84 | -1.0 | 252 |
| $\gamma$ Ser | -171 ± 89 | -1.9 | 267 |
| $\iota$ Peg | -169 ± 111 | -1.5 | 333 |
| $\iota$ Per | -281 ± 139 | -2.0 | 417 |
| $\iota$ Psc | -84 ± 106 | -0.8 | 318 |
| $\kappa$-1 Cet | -115 ± 172 | -0.7 | 516 |
| KX Lib | 469 ± 341 | 1.4 | 1492 |
| $\lambda$ Aur | 368 ± 190 | 1.9 | 938 |
| *NSV* 4765 | -564 ± 262 | -2.2 | 786 |
| $\tau$ Boo | 151 ± 101 | 1.5 | 454 |
| $\theta$ Per | -54 ± 111 | -0.5 | 333 |
| $\upsilon$ And | -72 ± 166 | -0.4 | 498 |
| | | Average: | 567 |

Table 6. Spitzer/IRS – KIN comparison.

| Name | HD | Spitzer/IRS [a] $\frac{Fdust}{F\star}$ | $z$ [b] (3$\sigma$ limit) | KIN $z$ (3$\sigma$ limit) |
|---|---|---|---|---|
| 47 Uma | 95128 | -0.022 $\pm$ 0.013 | 1110 | 628 |
| $\beta$ Com | 114719 | 0.014 $\pm$ 0.01 | 830 | 972 |
| $\gamma$ Lep | 38393 | 0.001 $\pm$ 0.01 | 750 | 252 |
| $\iota$ Psc | 2223658 | -0.007 $\pm$ 0.014 | 970 | 318 |
| kx Lib | 131977 | 0.002 $\pm$ 0.01 | 1600 | 1492 |
| $\tau$ Boo | 120136 | 0.011 $\pm$ 0.014 | 970 | 454 |
| $\theta$ Per | 16895 | 0.003 $\pm$ 0.01 | 750 | 333 |
| $\upsilon$ And | 9826 | -0.003 $\pm$ 0.01 | 970 | 498 |

[a]from Beichman et al. (2006a) and Lawler et al. (2009).

[b]computed as: $L_{dust}/L_\star \times 10^{-7}$.